\documentclass[%
 reprint,
 amsmath,amssymb,
 aps,
]{revtex4-2}

\usepackage{float}
\usepackage{xcolor}
\usepackage{amsmath}
\usepackage{amssymb}
\usepackage{hyperref} 
\usepackage{graphicx}
\usepackage{dcolumn}
\usepackage{bm}

\begin{document}


\title{Tunable magnonic crystal in a hybrid superconductor--ferrimagnet nanostructure}

\author{Julia Kharlan$^{1,2}$, Krzysztof Szulc$^{1,3}$, Jarosław W. Kłos$^{1}$,  Grzegorz Centała$^1$ }
\email{grzcen@amu.edu.pl}
\affiliation{
$^{1}$ISQI, Faculty of Physics, Adam Mickiewicz University, Poznań, Poland\\
$^{2}$Institute of Magnetism NASU and MESU, Kyiv, Ukraine\\
$^{3}${Institute of Molecular Physics, Polish Academy of Sciences, Poznań, Poland}
}%

\date{\today}

\begin{abstract}
One of the most intriguing properties of magnonic systems is their reconfigurability, where an external magnetic field alters the static magnetic configuration to influence magnetization dynamics. In this paper, we present an alternative approach to tunable magnonic systems. We studied theoretically and numerically a magnonic crystal induced within a uniform magnetic layer by a periodic magnetic field pattern created by the sequence of superconducting strips. We showed that the spin-wave spectrum can be tuned by the inhomogeneous stray field of the superconductor in response to a small uniform external magnetic field. Additionally, we demonstrated that modifying the width of superconducting strips and separation between them leads to the changes in the internal field which are unprecedented in conventional magnonic structures. The paper presents the results of semi-analytical calculations for realistic structures, which are verified by finite-element method computations.
\end{abstract}

\maketitle


\section{Introduction \label{sec:Intro}}

Magnonic crystals (MCs) \cite{Krawczyk_2014} are artificial magnetic materials, whose magnetic parameters are characterized by periodic variation in space.

Modulations of the magnetic properties in MCs can be achieved by periodic variation of such parameters as saturation magnetization, exchange constant \cite{Klos_2012}, film thickness \cite{Langer_2019}, magnetic anisotropy \cite{Choudhury_2020}, etc. Using the magnetization textures \cite{Yu_2021,Szulc2022}, including the creation of skyrmion-based MCs \cite{Muhlbauer_2009,Mruczkiewicz_2016}, has been the focus of recent investigations. Despite numerous studies performed in this research field, magnonic crystals are still an active topic \cite{Chumak_2017,Flebus_Roadmap_2024} because the periodic pattering of magnetic material is one of the easiest ways to tailor the dispersion relation of spin waves (SWs) and affect their localization and propagation.

One of the advantages of magnonic systems over their photonic or phononic counterparts is the relative ease with which reconfigurability can be introduced. The magnetic configuration, and thus the influence of magnetization dynamics, can be controlled with external stimuli. However, the process of magnetization switching in periodic structures \cite{Grundler_2010, Szulc_2019,Gartside_2021} and periodic magnetic textures \cite{Szulc2022,Gruszecki2017} is a nonlinear process, often irreversible or, in the best case, hysteretic. Another approach is the control of material parameters (such as perpendicular magnetic anisotropy) by an external (electric) field \cite{Choudhury_2020,Wang_2017}. This technique, however, is restricted to thin magnetic films and the range of linear changes of anisotropy is limited \cite{Rana_2018}.

In our theoretical and numerical work, we propose an alternative solution to induce the MC on demand and tailor SW propagation by a homogeneous, static external magnetic field. We use the eddy currents in a periodic sequence of superconducting (SC) strips in the Meissner state to create the profile of a static field inside the magnetic layer in response to the external field. The depth of the periodic field profile varies linearly with the external field which determines the strength of SW scattering and the width of magnonic gaps. 

The considered structure is an electromagnetically coupled superconductor--ferrimagnet hybrid. 

There are reports about the interactions in such hybrid systems between magnetic textures (domains, skyrmions) and SC planar structures (layers or disks) \cite{Dahir_2020, Gonzalez_2022}, where the inhomogeneous pattern of Meissner currents is induced in the SC layer by the stray field of non-uniform magnetization configuration. It is also possible that the stray field produced by Meissner currents in SC nanoelement (planar dot) induces the magnetization texture in the magnetic subsystem (layer) \cite{Palau_2016}. Although this interaction is mutual in principle, its effects (change of magnetic configuration and induction of eddy currents) depend on the specific realization of the hybrid system, and in particular on its geometry.

The coupling affecting the magnetization dynamics \cite{Golovchanskiy2018,borst_2023,Kharlan2024} is, in general, both static and dynamic. In the first case, the static field produced by the superconductor acts as the non-uniform internal field which can modify the orientation of equilibrium magnetization. The presence of magnetic texture and non-uniform internal field influences spin-wave localization and propagation. The latter scenario corresponds to the direct influence of the dynamic stray field of the superconductor on SWs.

The superconductor--ferrimagnet hybrids can also be periodic structures. The periodicity can be introduced in a top-down process, i.e. by the pattering of pristine materials for magnetic and/or SC component\cite{Milosevic_2004, Golovchanskiy2018, Khaydukov_2019}. However, both components can exhibit, in a bottom-up processes, the intrinsic periodicity related to the presence of the periodic magnetization texture or the lattice of vortices of supercurrent\cite{Dahir_2020, Dobrovolskiy_2019, Jafri_2020}, which are nucleated under certain conditions (applied field, material parameters, temperature). 
The introduction of periodicity creates a magnonic crystal for the SWs propagating in the magnetic component of the hybrid structure. Particularly interesting is the case when the periodicity is introduced only in the SC component and then perceived in the uniform magnetic part due to coupling. This approach allows to tune the properties of the magnonic system indirectly, thus extending its controllability with respect to conventional magnonic crystals.

As we can see, periodic superconductor--ferrimagnet hybrids can be realized in many ways and their operating mechanisms can be different. For this reason, their theoretical description is generally very complex. Therefore, we have proposed a system that is relatively easy to describe semi-analytically and, on the other hand, realizes the function of a controllable MC. The studied system (shown in Fig.~1) has the following features: (i) the periodicity is permanent only in the SC subsystem, and the coupling, tuned by external bias, induces the periodicity in the magnetic subsystem -- such features demonstrate the unconventional controllability of the MC,
(ii) the dynamic coupling is small, and the static coupling affects the internal magnetic field without significant change of the magnetization configuration, (iii) the SC system remains in the Meissner state, (iv) the SC and magnetic components are not in direct contact \cite{Khaydukov_2019,Putilov_2022}, (v) the London penetration depth $\lambda$ is finite. The properties (ii)-(iv) are obtained by appropriate design of the system and by using low values of the applied field. They greatly simplify the theoretical modeling of the system. Point (v) generalizes the theoretical approach used in the pioneering experimental works of Golovchanskiy \emph{et al.} \cite{Golovchanskiy2018,Golovchanskiy2019,Golovchanskiy2020a}, based on the assumption of ideal diamagnetism of the superconductor ($\lambda=0$). 

This paper is organized as follows. In the section 'Structure', after the introduction, we describe the studied superconductor--ferrimagnet hybrid system. In the latter section 'Results', we present the outcomes showing the stray field produced by the SC pattern and the SW spectra in the FM layer. The results are summarized in the 'Discussion' section. In the subsequent 'Methods' section, we describe semi-analytical models concerning the field produced by a periodic system of SC strips in the Meissner state and the determination of the SW spectrum of a ferrimagnetic (FM) layer placed in a periodic external field. 

\section{Structure}

\begin{figure}[t!]
\centering
\includegraphics[width=0.95\columnwidth]{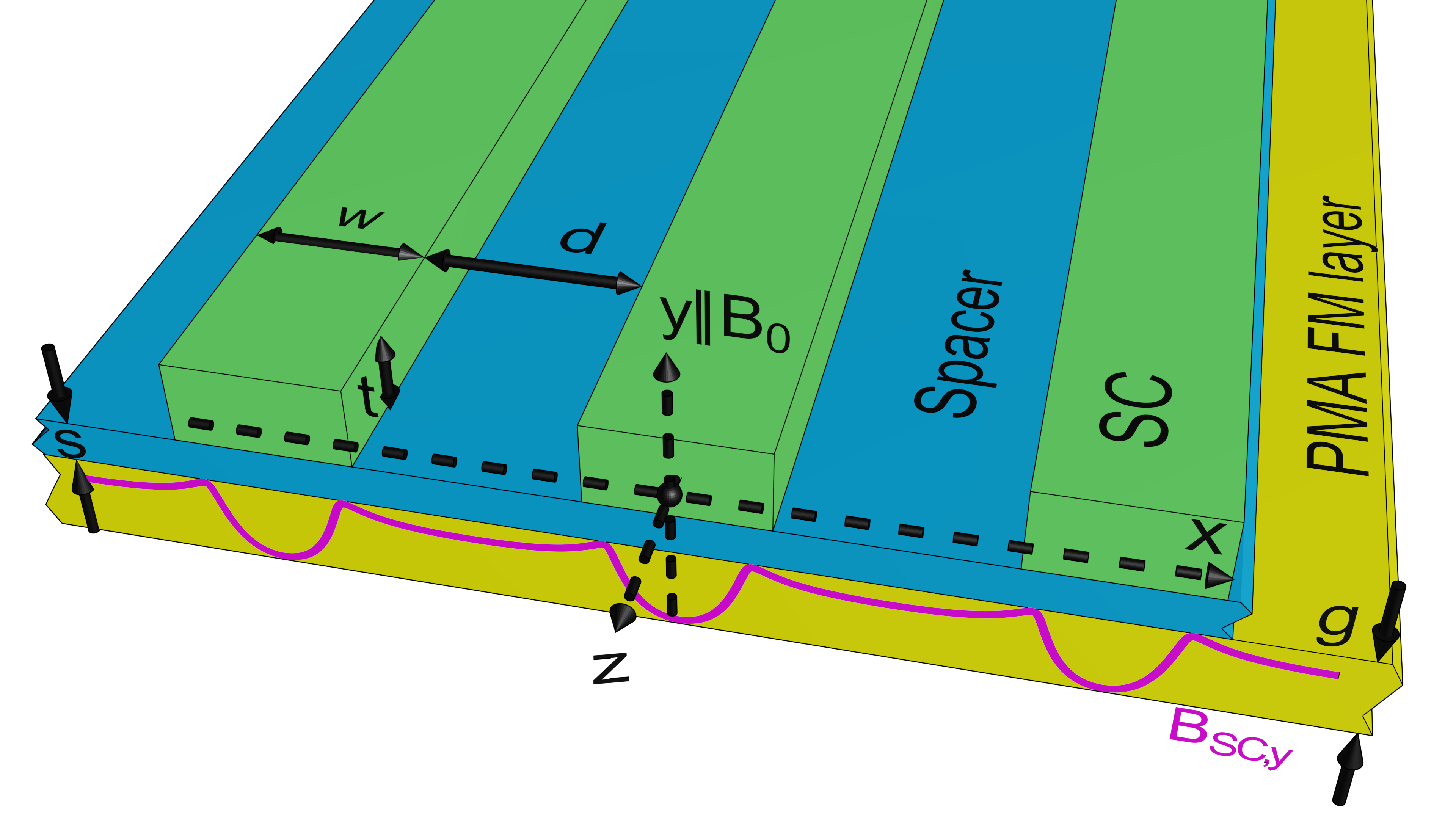}
\caption{A thin FM film ($g=20$ nm) is placed in the periodic stray field of the sequence of SC strips ($w=400$ nm, $t=100$ nm). The film, made of Ga:YIG, and the SC strips, made of Nb, are separated by a nonmagnetic spacer ($s=10$ nm). Periodic stray field (magenta line) is responsible for the formation of magnonic crystal. We study the impact of the external field $B_{0}$ and the width of the gaps $d$ between SC strips on the SW spectrum in magnonic crystal.}
\label{fig:structure}
\end{figure}

We investigated theoretically and numerically the SW dynamics in a realistic superconductor--ferrimagnet hybrid, which consists of gallium-doped yttrium iron garnet (Ga:YIG) ferrimagnetic (FM) film and the sequence of Nb SC strips in Meissner state. Both SC and FM subsystems are electrically isolated from each other by a 10~nm-thick nonmagnetic spacer (see Fig.~\ref{fig:structure}). This geometry only allows for electromagnetic coupling between the subsystems, thus avoiding any proximity effects at the ferrimagnet--superconductor interface.
The system is placed in the external magnetic field perpendicular to the film plane.    The 20~nm-thick FM film is characterized by the following values of material parameters: saturation magnetization $M_{\rm s}=16$~kA/m, exchange stiffness $A_\mathrm{ex}=1.37$~pJ/m and gyromagnetic ratio $\gamma=179$~rad/(T$\,$ns). The Ga:YIG film has strong perpendicular magnetic anisotropy (PMA) \cite{Bottcher2022}, which surpasses the shape anisotropy and leads to the out-of-plane magnetization ground state even in the absence of external magnetic field. We assumed the realistic value of the uniaxial anisotropy $K_\mathrm{u}=756$~J/$\mathrm{m}^3$~\cite{Bottcher2022}. The out-of-plane orientation of magnetization remains stable in the external field applied in the same direction. For such a configuration, the SC strips will produce the wells of stray field in the FM layer.  Moreover, this configuration minimizes the dynamic coupling between the SC strips and the FM layer\cite{Kharlan2019}. The other important property of Ga:YIG is its low SW damping, which is essential for magnonic applications. We selected the widely used SC material -- Nb. The Nb strips (400 nm wide, 100 nm thick) are characterized by the London penetration depth $\lambda = 50$~nm \cite{Gubin2005} which ensures the relatively strong screening. The Meissner currents in SC strips create periodic magnetic field distribution inside the FM film and, therefore, transform homogeneous film into one-dimensional magnonic crystal (MC). 

In this study, similarly as in our previous work \cite{Kharlan2024}, the impact of the FM layer on the dynamics of SC currents can be neglected and the SWs can be studied in the static stray field generated by the SC structure.  Therefore, this problem can be solved in two steps. First, the stray field generated by the periodic system of SC strips is determined by solving the London equation (LE), using the approach proposed by Brandt \cite{Brandt1994a, Brandt1994b}, and then the spectrum of SWs in this field is calculated by solving the Landau--Lifshitz (LL) equation, using the plane-wave method (PWM) \cite{Krawczyk_2012}. These semi-analytical calculations are further compared with the numerical solution of the dynamically uncoupled London and LL equations, using the finite-element method (FEM) simulations in COMSOL Multiphysics. In our system, the absence of dynamic coupling has a negligible effect on the results of the FEM simulations, as discussed in the Supplementary Informations, section 3 (SI~3).

\section{Results}

\subsection*{Stray field produced by the sequence of superconducting strips}

\begin{figure}[!h]
\centering
\includegraphics[width=0.95\columnwidth]{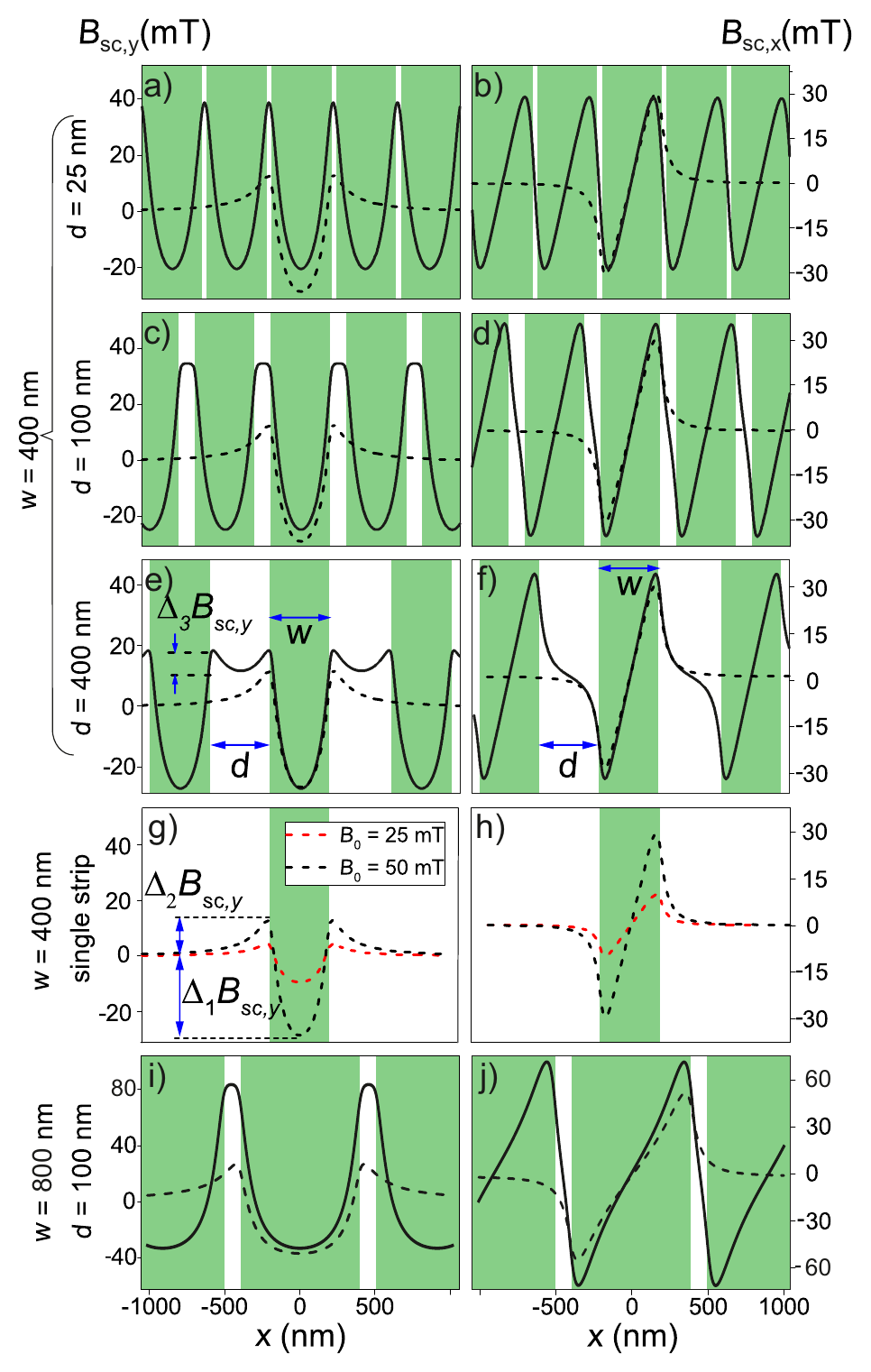}
\caption{The profiles of the magnetic field components $B_{{\rm sc},y}$ (a,c,e) and $B_{{\rm sc},x}$ (b,d,f) produced by the sequence of SC strips placed in the external magnetic field $B_{0}=50$~mT for the distance between adjacent strips $d=25$~nm (a,b), $d=100$~nm (c,d), $d=400$~nm (e,f). For comparison, the profiles of $y$- and $x$-components of stray field generated by a single SC strip are shown in (a-f) by dashed lines. The depth of $\mathbf{B}_{\rm sc}(x)$ profile is linearly increased with $B_{0}$ -- see e.g. (g,h) where the red and black dashed lines correspond to the cases of external field $B_{0}=25$~mT and 50 mT for single strip, respectively. While keeping the same spacing ($d=100$~nm) and external field ($B_0=50$~mT), the widening of the SC strips to $w=800$~nm results in an increase in the height of the barriers separating the wells, and the increase of the in-plane component $B_{{\rm sc},x}$ tilting the static magnetization - compare (i,j) with (c,d). All profiles are plotted along the line $y=-70$~nm passing through the center of the FM film.}
\label{fig:SC field}
\end{figure}

Figure~\ref{fig:SC field} shows the $y$-component (Fig.~\ref{fig:SC field}(a,c,e)) and the $x$-component (Fig.~\ref{fig:SC field}(b,d,f)) of the stray field $\mathbf{B}_{\rm sc}$ generated by the sequence of SC strips in response to an out-of-plane external magnetic field $\mathbf{B}_{0}=B_0\hat{\mathbf{y}}$ ($B_0=50$~mT). All profiles are calculated in the middle of the FM film, i.e. for $y=-70$~nm, because the changes of $B_{\rm sc}(y)$ over the thin FM layer are not significant. We considered three selected values of the distance between adjacent strips: $d=25$~nm (Fig.~\ref{fig:SC field}(a,b)), $d=100$~nm (Fig.~\ref{fig:SC field}(c,d)), $d=400$~nm (Fig.~\ref{fig:SC field}(e,f)), for fixed strip width $w=400$~nm, to study the effect of the distance $d$ on the field of the strips $\mathbf{B}_{\rm sc}(x)$. In Fig.~\ref{fig:SC field}(g,h) we have shown both components of the stray field: $B_{{\rm sc},y}(x)$ and $B_{{\rm sc},x}(x)$, generated by a single SC strip for reference, i.e. for the case $d\rightarrow\infty$ in which the strips do not interact. The red and black dashed lines in Fig.~\ref{fig:SC field}(g,h) correspond to the cases of the external field $B_0=25$~mT and 50~mT, respectively. Since the stray field is proportional to the applied field (see Eqs.~(\ref{eq:curr_int2}) and~(\ref{eq:b_field})), the profiles Fig.~\ref{fig:SC field}(g-h) scale linearly as $B_0$ increases from 25~mT to 50~mT -- compare the changes in depths of the wells $\Delta_1 B_{{\rm sc},y}$ and heights of the barriers $\Delta_2 B_{{\rm sc},y}$. Therefore, we have skipped the plots $\mathbf{B}_{\rm sc}(x)$ for $B_0=25$~mT in Fig.~\ref{fig:SC field}(a-f). The black dashed lines in Fig.~\ref{fig:SC field}(a-f) show, for reference, the profiles of field produces by since SC strip for $B_0=50$~mT. In the last row (Fig.~\ref{fig:SC field}(i,j)), we present the results for the sequence of wider strips $w=800$~nm, to compare them with the  case of narrower strips $w=400$~nm (placed in the same field $B_0=50$~mT and separated by the same gaps $d=100$~nm as in Fig.~\ref{fig:SC field}(c,d)), and discuss the interplay between the geometrical parameters $w$, $d$ and external field $B_0$ which is important for the formation of the barriers $\Delta_2 B_{{\rm sc},y}$.

Due to the Meissner effect, the SC strips expels the magnetic field from its interior (green areas in Fig.~\ref{fig:SC field}). This effect is achieved by the induction of eddy currents, which create the internal field well to compensate the applied field. The lines of the total magnetic field are then simply pulled out of the SC strips and form an onion-like shape where the line density is reduced inside and increased outside the SC strip body. This means that the field $B_{{\rm sc},y}$ produced by the superconductor forms the barriers in the gaps between them (see white areas in Fig.~\ref{fig:SC field}(a,c,e)). Moreover, the deflection of the field lines of the superconductor is equivalent to the induction of the asymmetric component of the stray field $B_{{\rm sc},x}$, perpendicular to the external field $\mathbf{B}_0=B_0\hat{\mathbf{y}}$ (right column in Fig.~\ref{fig:SC field}). 

\begin{figure*}[t]
\centering
\includegraphics[width=0.95\textwidth]{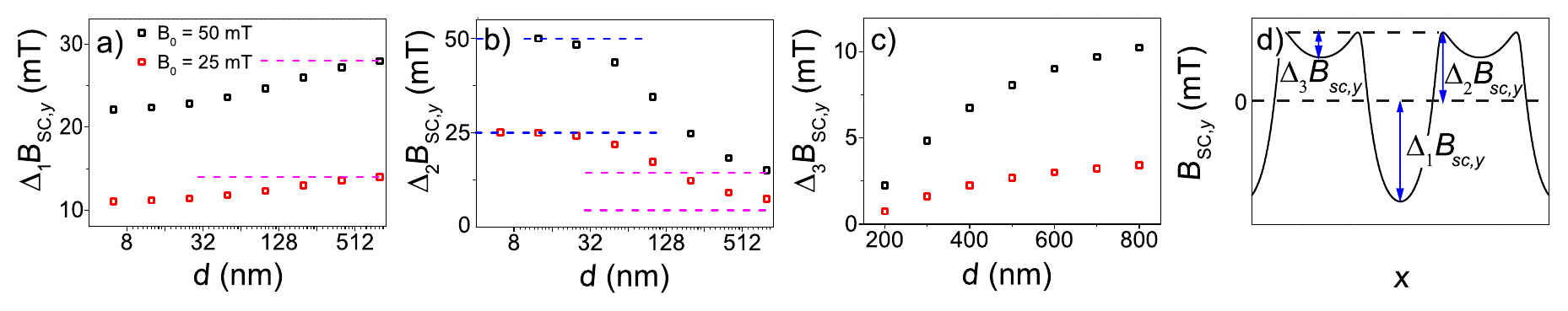}
\caption{The changes of the depth of the wells (under SC strips) $\Delta_{1}B_{{\rm sc},y}$ (a), the height of the barriers (close to the edges of strips) $\Delta_{2}B_{{\rm sc}, y}$ (b), and the shallower wells (in the gaps between the strips) $\Delta_{3}B_{{\rm sc},y}$ (c) with the the gaps $d$. The profile of the out-of-plane component of the stray field of the SC strips $B_{{\rm sc},y}(x)$, for which $\Delta_{1}B_{{\rm sc},y}$, $\Delta_{2}B_{{\rm sc},y}$, $\Delta_{3}B_{{\rm sc},y}$ are determined (d), is taken in the center of the FM layer. The blue and magenta lines show the asymptotic behavior for $d\rightarrow 0$ and $d\rightarrow \infty$, respectively. The lowest value of $d\approx 150$~nm in (c) corresponds to the case when minima between the barriers disappear. The values $\Delta_{1}B_{{\rm sc},y}$, $\Delta_{2}B_{{\rm sc},y}$, $\Delta_{3}B_{{\rm sc},y}$ change linearly with the external field $B_0$ -- we presented their values for: $B_0=25$~mT (black points) and 50~mT (red points).}
\label{fig:delta B}
\end{figure*}

Reducing the distance between the SC strips leads to their mutual interaction and modification of the $\mathbf{B}_{\rm sc}(x)$ profile. It is easy to see from Fig.~\ref{fig:SC field}(a,c,e) that decreasing the distance $d$ between the strips leads to a significant increase in the barrier height $\Delta_{2}B_{{\rm sc},y}$. This interesting effect can be explained intuitively by noting that the field lines expelled from the SC strips (green areas) must be concentrated in narrow regions separating the strips (white areas). In addition, the width of the SC strips is also relevant for barrier height. The wider strips must expel more flux from their interior, therefore, the concentration of the larger amount of field lines in the gaps of the same size must lead to an increase of the barrier height -- compare Fig.~\ref{fig:SC field}(c) and Fig.~\ref{fig:SC field}(i). Assuming ideal diamagnetism of the SC strips (zeroing of the London penetration depth: $\lambda=0$), i.e. considering that all the magnetic flux from their interiors is expelled into the gaps between them, we can estimate the average  height of the barrier inside the gaps: $\langle B_{{\rm sc},y}\rangle\approx B_0\, w/d$.

It is interesting to note that for relatively large distances $d$, the minimum  appear at the top of each barrier of $B_{{\rm sc},y}$ (Fig.~\ref{fig:SC field}(e)). These minima disappear as the SC strips come closer together and transform into a sharp peak (Fig.~\ref{fig:SC field}(a)) through an intermediate small plateau (Fig.~\ref{fig:SC field}(c)). 

With decreasing distance $d$ between the SC strips, the depth of the stray field wells $\Delta_{1}B_{{\rm sc},y}$ does not change as dramatically as the height of the barriers $\Delta_{2}B_{{\rm sc},y}$.  The wells become slightly shallower as the SC strips begin to feel each other. This effect is related to the increase in the height of the barriers for small $d$ and the continuity of the field on the side faces of the SC strips. As an estimation (strict for layer extended in $yz$-plane \cite{Tinkham_2004}), the field inside the strip decays exponentially: $B\approx B_{\rm edge}\cosh((x-x_0)/\lambda)/\cosh(w/(2\lambda))$. Therefore, for larger values of the field on the lateral faces of the strip $B_{\rm edge}$, the minimum of the field is shallower in the center $x=x_0$.

It is worth noting that the in-plane component of the stray field $B_{{\rm sc},x}$ does not change significantly with increasing distance $d$ between SC strips -- see the right column in Fig.~\ref{fig:SC field}. The $B_{{\rm sc},x}$ is responsible for tilting the static magnetization and reaches extreme values in narrow regions near the strip edges. Therefore, the magnetization texture induced in FM will remain relatively weak regardless of $d$ and its change will only be related to the modification of the period of the SC pattern and the different locations of the strip edges.

The dependence of the well depth $\Delta_{1}B_{{\rm sc},y}$ and the barrier height $\Delta_{2}B_{{\rm sc},y}$ on the SC strip separation is shown in Fig.~\ref{fig:delta B}(a,b). The well depth $\Delta_{1}B_{{\rm sc},y}$ is relatively insensitive to the strip spacing and decreases only slightly with decreasing $d$. Clearly, $\Delta_{1}B_{{\rm sc},y}$ reaches the depth of the well of a single SC strip in the $d\rightarrow \infty$ limit (dashed magenta lines). As discussed earlier, the barrier height $\Delta_{2}B_{{\rm sc},y}$ increases for $d\rightarrow 0$. However, the numerical studies show that this increase is not unlimited. The maximum (dashed blue lines) is higher for smaller values of $d/w$, i.e., when the flux expelled from wider SC has to concentrate in narrower gaps and is linearly scaled with the value of the applied field $B_0$. As expected, $\Delta_{2}B_{{\rm sc},y}$ decreases to the barrier height of the single strip (dashed magenta lines) for large $d$. It is clear that the total effective well depth $\Delta B_{{\rm sc},y}=\Delta_{1}B_{{\rm sc},y}+\Delta_{2}B_{{\rm sc},y}$ increases with the reduction of the gap $d$. 

Figure~\ref{fig:delta B}(c) shows the evolution of the well between the barriers in the gap $\Delta_{3}B_{{\rm sc},y}$ with the increase of $d$. The well appears only for larger distances between the strips (in our case for $d>200$~nm, for the strip of width $w=400$~nm). For smaller values of $d$ the barriers merge and the minimum disappears. 

The stray field generated by a periodic sequence of SC strips can create a periodic landscape of the internal field in the uniform magnetic layer, thereby inducing a magnonic crystal on demand, e.g. by the lowering of temperature when the Meissner state in SC strips is induced. The depth of this landscape can be tailored by adjusting the spacing between the strips and then tuned linearly by an external magnetic field.

\subsection*{Spin-wave dispersion in magnonic crystal induced by a superconducting pattern}

Let us study the magnetization dynamics in a uniform FM layer with out-of-plane anisotropy, where the magnetization precesses around the normal to the layer, i.e. in the $y$-direction -- see Fig.~\ref{fig:structure}. If the layer is placed in a stray field created by a periodic sequence of SC strips (with an external field applied perpendicular to their plane), it becomes a magnonic crystal. The SWs scatter in a periodic field and their eigenfunctions are Bloch waves, which propagate only in certain frequency ranges, called allowed frequency bands. 

\begin{figure*}[!t]
\centering
    \includegraphics[width=0.95\textwidth]{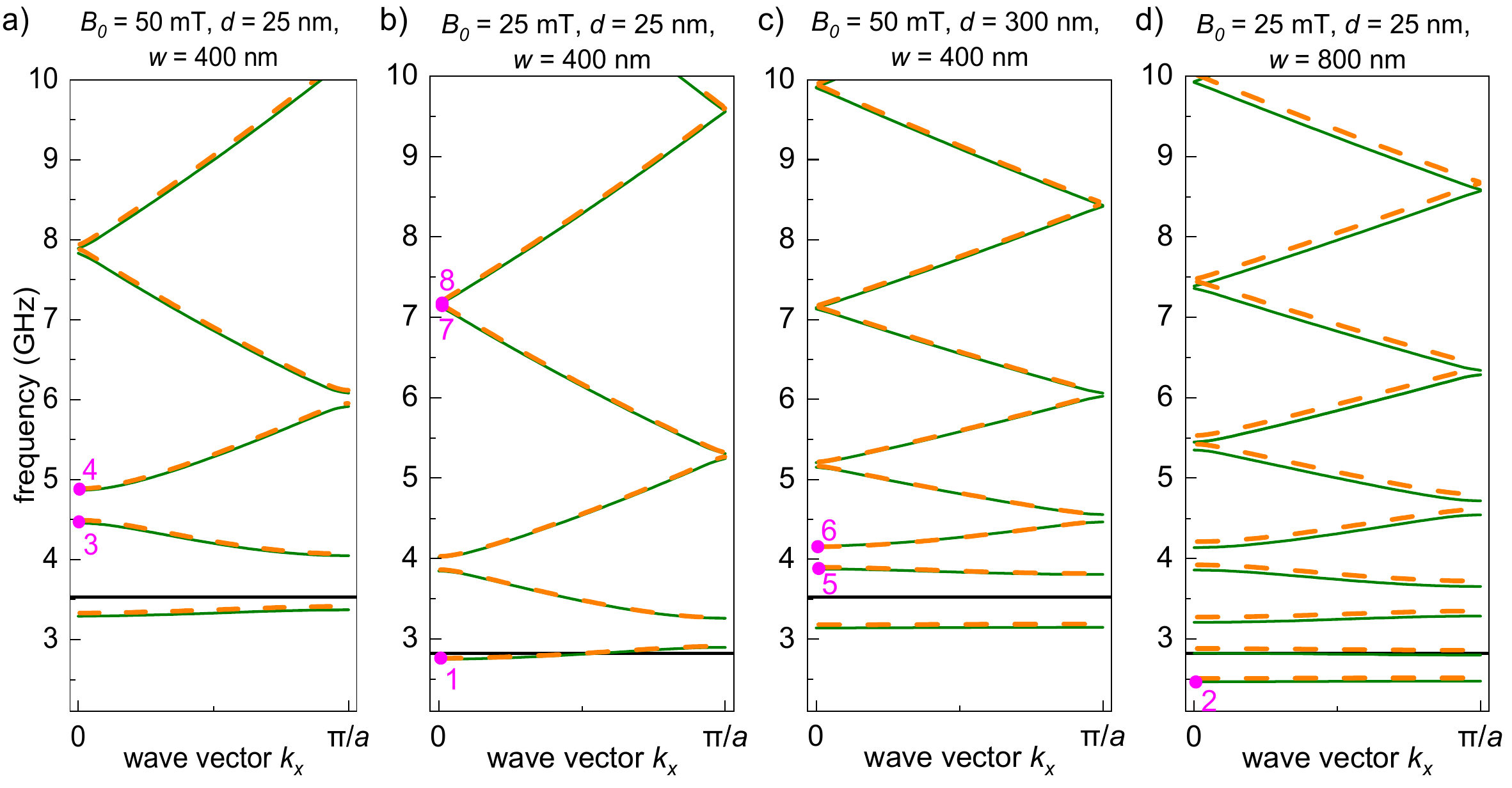}
    \caption{The SW dispersion relations in the FM layer under the influence of the stray field produced by the sequence of SC strips -- Fig.~\ref{fig:SC field}. We have changed (b) the external field, (c) the width of the gaps, and (d) the width of the strips and the external field, with respect to the reference system (a). The dashed orange (solid green) lines represent the results of PWM (FEM) calculations. The horizontal black solid lines indicate the FMR frequency for the pristine Ga:YIG layer. The magenta labels 1-8 indicate the bands and wave vector for the selected SW profiles plotted in Fig.~\ref{fig:Profiles}. }
\label{fig:Dispersion_relations}
\end{figure*}

The exemplary dispersion relations for the SWs propagating in the magnonic crystal described above are shown in Fig.~\ref{fig:Dispersion_relations}. The dispersion presented in Fig.~\ref{fig:Dispersion_relations}(a) is the reference one. One of the key parameters is the external field $B_{0}$. In our study, we use a maximum of 50~mT because higher fields result in a mixed state of the SC strip \cite{Kharlan2024}. By lowering the external field for the same geometry (compare Fig.~\ref{fig:Dispersion_relations}(a) and Fig.~\ref{fig:Dispersion_relations}(b)), we can lower the frequencies of all bands. However, in the considered hybrid system, the applied field $B_0$ is also responsible for generating the periodic landscape of the internal magnetic field in the FM layer. The weaker $B_0$ is, the smaller the stray field generated by the superconductor. This leads to the shallowing of the wells (see red and black dashed lines in Fig.~\ref{fig:SC field}(g)) and to a reduction of the SW scattering over the wells. These effects are clearly reflected in Fig.~\ref{fig:Dispersion_relations}(b), where (i) the lowest magnonic band is almost pushed out of the shallow well -- see the position of this band with respect to the FMR frequency of the pristine layer (solid black line in Fig.~\ref{fig:Dispersion_relations}) and (ii) the magnonic gaps are significantly narrowed as a result of weaker scattering. If we keep the same value of the external field $B_0=50$~mT, but increase the distance between the SC strips to $d=300$~nm (Fig.~\ref{fig:Dispersion_relations}(c)), the magnonic gaps are also narrowed, referring to the case shown in Fig.~\ref{fig:Dispersion_relations}(a). This can be understood if we notice that the scattering will be weaker in the system where the wells occupy the smaller fraction of space in FM layer. The lowered frequency of the first band results from the deepening of the stray-field well -- see Fig.~\ref{fig:delta B}.
If we keep the field low ($B_0=25$~mT), we can still push the bands inside the wells by widening the SC strips, and the corresponding stray-field wells will be widened as well. In Fig.~\ref{fig:Dispersion_relations}(d), where $w=800$~nm, we can see that the wells can accommodate two bands. 

In Fig.~\ref{fig:Dispersion_relations}(c,d) the periods: $a=w+d$ are extended due to the increase of $w$ or $d$. This reduces the size of the first Brillouin zone and leads to more folds of the dispersion relation, which generates more magnonic bands. As a result, the bands in Fig.~\ref{fig:Dispersion_relations}(c,d) become narrower.

To calculate the dispersion relations, we used two different approaches. The solid green line in Fig.~\ref{fig:Dispersion_relations} presents the results of the FEM computations performed in COMSOL Multiphysics environment. In these calculations, we solved the linearized LL equation for a weak magnetization texture induced in the presence of the non-uniform stray field of the SC strips and by considering a small change of this field through the thickness of the FM layer. To demonstrate that neither the slight tilt of the equilibrium magnetization nor the inhomogeneity of the field significantly affects the SW spectrum, we performed the calculations based on the PWM assuming the homogeneity of the field across the thickness of FM layer and the uniform out-of-plane oriented static magnetization. The results of the semi-analytical PWM (dashed orange line) and the FEM (solid green line) are very similar and show small discrepancies, which are noticeable only for the case of wide SC strips - Fig.~\ref{fig:Dispersion_relations}(d).  We assumed that there is no dynamic coupling between the  magnetization in FM layer and eddy currents in SC strips \cite{Kharlan2024}. However, to demonstrate that we are justified in making this assumption for the considered system, we performed the supplementary FEM simulation, presented in SI~3. We considered there the limit of ideal diamagnet for SC strips, which means the perfect shielding ($\lambda=0$), i.e. complete expulsion of magnetic field, including dynamic one, from their interiors. We barely observe any changes in the SW frequencies -- see inset in Fig.~S1 in Supplementary Informations.

\begin{figure}[!b]
\centering
    \includegraphics[width=0.95\columnwidth]{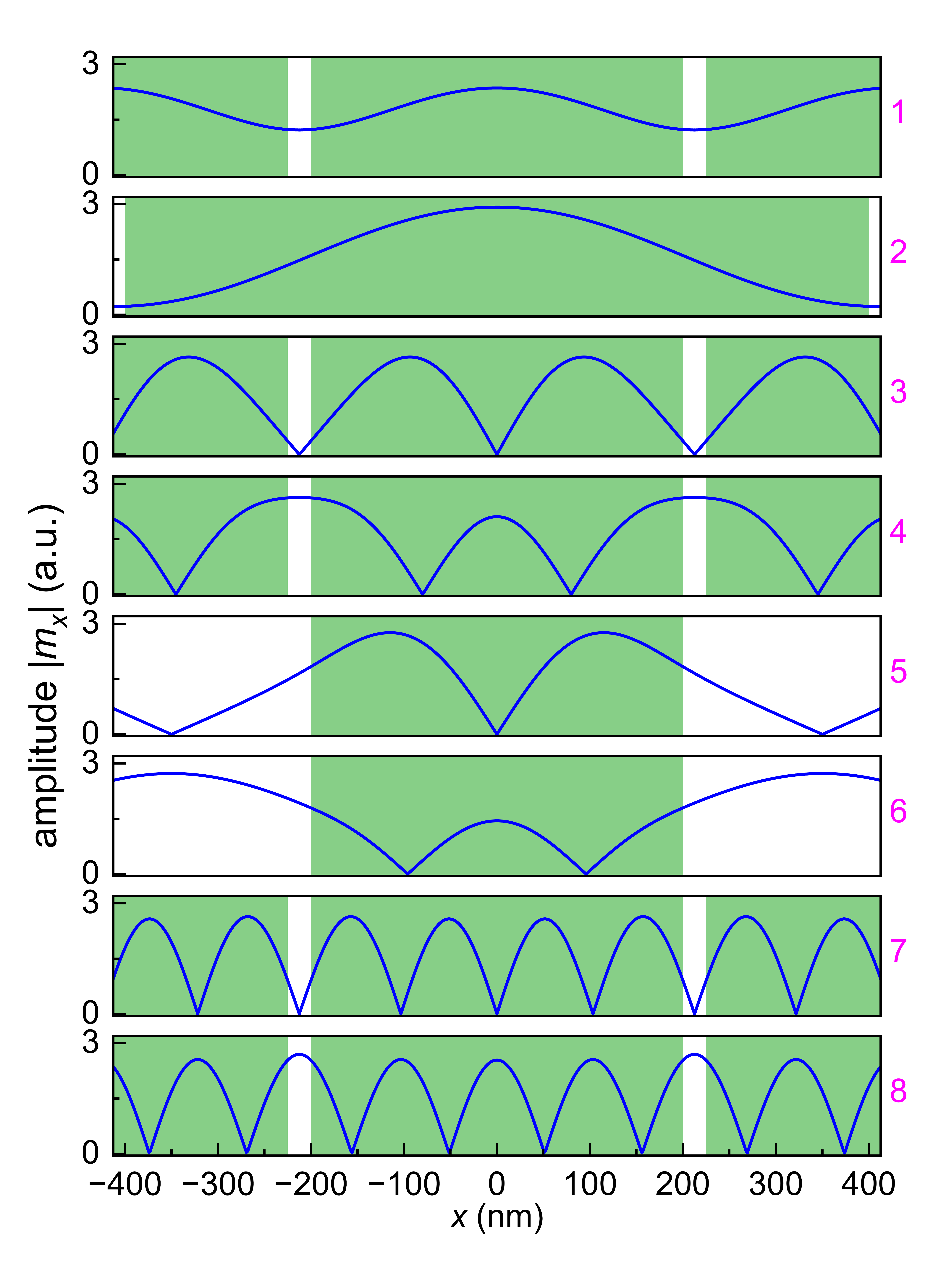}
    \caption{The SW profiles $|m_{x}|$ at $k_x=0$ for selected bands -- compare the magenta numbers on the right with those in Fig.~\ref{fig:Dispersion_relations}. The profiles are nearly uniform across the layer and were taken in the center of the layer. The green (white) areas indicate the position of the SC strips above the FM layer (the gaps between the SC strips).}
\label{fig:Profiles}
\end{figure}

The magenta labels in Fig.~\ref{fig:Dispersion_relations}, mark the bands for which we have shown the SW profiles in Fig.~\ref{fig:Profiles} at $k_x=0$. The profiles are plotted in the center of the FM layer. However, this choice is arbitrary, because for the considered thin FM layer (20 nm) all modes are almost uniform over the layer's thickness for all investigated bands (for the frequencies below 10 GHz). Let's discuss the spatial distribution of the in-plane component of the dynamic magnetization $|m_x|$ for the modes shown in Fig.~\ref{fig:Dispersion_relations}. The profiles of the modes labeled 1 and 2 are shown in the top two rows of Fig.~\ref{fig:Profiles}. They represent the first narrow bands. These modes are confined in the wells of the stray field of the SC strips and their amplitude is reduced in the areas under the gaps between the strips (white areas in Fig.~\ref{fig:Profiles}). This reduction is stronger for the modes confined in wider strips ($w=800$~nm, mode labeled 8), where the barriers at the edges of the strips separating the wells are higher. The SWs in higher bands, above the FMR frequency of the pristine FM layer, propagate quite freely, since the SWs exhibit oscillatory behavior in both the regions below the strips and below the gaps. The modes within each of the pairs labeled (3,4), (5,6), and (7,8) are taken at the edges of a gap.
As expected, they are standing modes (one symmetric, another antisymmetric with respect to the center of the primitive cell), which have the same number of nodes. For a wide gap, where the frequencies of the modes in the mentioned pairs are noticeably different, the amplitudes in both components of the structure (here regions of barriers and wells, marked by green and white areas in 
Fig.~\ref{fig:Profiles}, respectively) are noticeably different. We see such an effect for the mode pairs labeled (3,4) and (5,6), i.e. for the edges of the gap between the first and fourth band, which are relatively wide (Fig.~\ref{fig:Dispersion_relations}(a,c)). The higher gaps are naturally narrower because the SWs of higher frequencies and shorter wavelengths scatter weaker on the real smooth landscape of the internal field. Therefore, the gap between the seventh and the eighth band is small (7,8) -- see Fig.~\ref{fig:Dispersion_relations}), and the profiles of the SWs at the edges of this gap resemble sine and cosine waves with practically constant amplitudes.

\begin{figure}[t]
\centering
    \includegraphics[width=0.95\columnwidth]{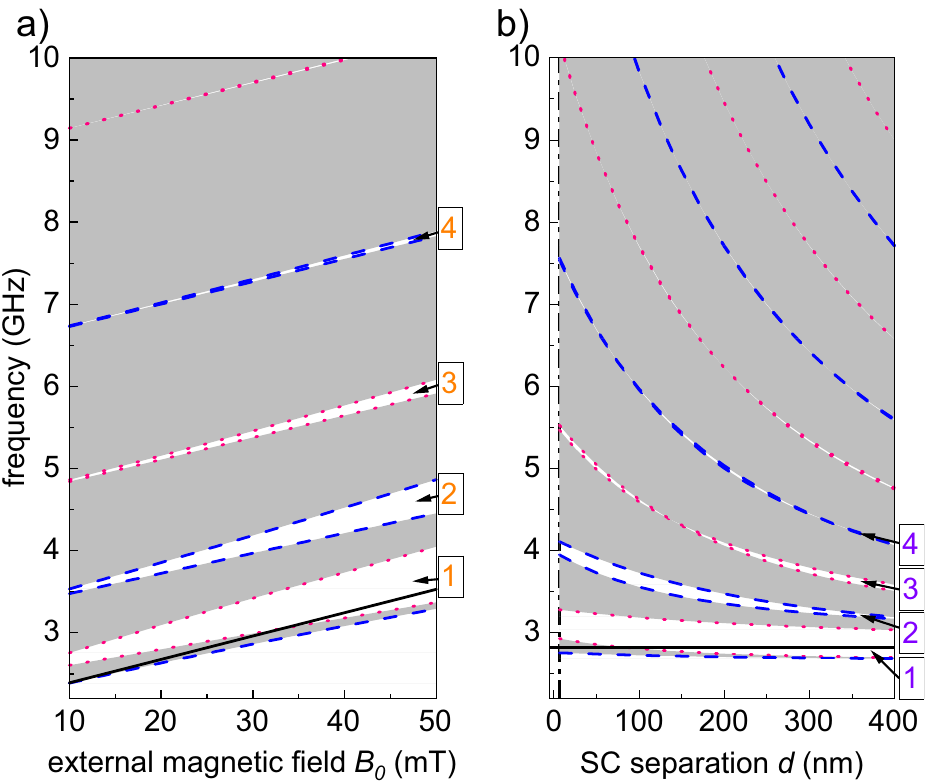}
    \caption{The SW spectrum as a function of (a) external magnetic field $B_0$ and (b) SC strip separation $d$. The frequency bands (gray regions) are bounded by the frequencies corresponding to the center of the 1st Brillouin zone $k_{x}=0$ (dashed blue line) and the edge of the 1st Brillouin zone $k_{x}=\pi/a$ (dotted pink line). The vertical black dash-dot line in (b) marks the minimum separation of SC strips (6.25~nm) for which numerical calculations were performed. Black solid lines denote the FMR frequency of the pristine FM layer, i.e. in the absence of SC pattern. Results presented in (a) and (b) were obtained for fixed $d=25$~nm and B$_{0}=25$~mT, respectively. The width of SC strips is set to $w=400$~nm both for (a) and (b).}
\label{fig:B0_sep}
\end{figure}

So far, we have demonstrated that it is possible to induce a magnonic crystal using a stray field generated by a periodic SC pattern. Let's discuss in more detail how we can tailor and control the SW spectrum on demand. In general, the spectrum is tailored by selecting the shape, size, and material parameters of the structure under consideration and can be controlled by an external field. 
In our study, we will show that changing the spacing between the elementary cells and the strength of the external magnetic field leads to changes in the SW spectrum that are not observable for conventional magnonic systems. Fig.~\ref{fig:B0_sep} shows the dependence of the SW spectrum on two parameters: the value of the external magnetic field $B_0$ (Fig.~\ref{fig:B0_sep}(a)) and the width of the gaps between the SC strips $d$ (Fig.~\ref{fig:B0_sep}(b)). The gray (white) areas indicate the frequency ranges corresponding to the frequency bands (gaps). For considered forward geometry ($\mathbf{k}\perp \mathbf{B}_0$), the edges of the bands for 1D magnonic crystal always appear in the center ($k_x=0$) or at the edge ($k_x=\pi/a$) of the Brillouin zone -- see Fig.~\ref{fig:Dispersion_relations}. Therefore, the edges of the bands have been plotted for the frequencies $f(k_x=0)$ with dashed lines and $f(k_x=\pi/a)$ with dotted lines. 

In both parts of Fig.~\ref{fig:B0_sep}, we can distinguish two regions. Narrow bands appear at frequencies lower than the FMR frequency of the pristine FM layer (solid black line in Fig.~\ref{fig:B0_sep}), while wider bands are found at higher frequencies. The range of frequency $\Delta f_1$ in which the narrow bands appear below the FMR frequency can be estimated from the depth of the stray field well $\Delta_1B_{{\rm sc},y}$  (see Fig.~\ref{fig:delta B}): $\Delta f_1 =\tfrac{1}{2\pi}|\gamma| \Delta_1B_{{\rm sc},y}$. We can see in Fig.~\ref{fig:delta B}(a) that as the external field $B_0$ increases and $\Delta_1B_{{\rm sc},y}$ grows, the frequency is pulled down. It is worth noting that the height of the barrier $\Delta_2B_{{\rm sc},y}$ similarly determines the range of frequencies (above the FMR frequency) in which the propagating SW must tunnel through the stray-field barriers. This is expected to reduce the SW group velocity and make the band(s) appearing just above the FMR frequency narrower. This effect is practically invisible in Fig.~\ref{fig:B0_sep}. For similar reasons, we cannot observe the effect of tunneling through the barriers around a small well $\Delta_3B_{{\rm sc},y}$ on the SW spectrum.  

Fig.~\ref{fig:B0_sep}(a) shows the tuning of SW spectrum by homogeneous external field $B_0$. It is well known that the increase of the uniform external magnetic field shifts the SW spectrum upwards. However, in the investigated hybrid structure, the height of the barriers and the depth of the wells of the stray field generated by the SC strips also increase with the external field. It implies that the SW scattering in the periodic landscape of the effective field is enhanced and the frequency gaps become wider with increasing $B_0$. Such effect is hardly achievable in conventional magnonic system where static effective field depends on magnetization configuration which changes nonlinearly with applied field (e.g. after reaching saturation the static exchange field is negligible and static demagnetizing field is fixed and determined by saturation magnetization and geometry of the sample). In the considered superconductor--ferrimagnet hybrid system, the main contribution to the periodic internal field comes from the stray field of the SC pattern, which changes linearly with the external field $B_0$ until it reaches the first critical field, when the Abrikosov vortices start nucleating. This linear effect can be related to the widening of the frequency gaps with the increase of the external field $B_0$ -- see Fig.~\ref{fig:B0_sep}(a). It was shown that, for bi-component magnonic crystals \cite{Tacchi2012}, the widths of the successive gaps are proportional to the consecutive Fourier coefficients material parameter (saturation magnetization). The similar arguments for the relation between Fourier coefficients of the models parameters, which define periodicity of the system (here static stray field), and the width of successive frequency gaps can be presented for considered hybrid magnonic crystal. The width of $n^{\rm th}$ gap almost proportional to $|H_{{\rm sc},G}|$, where $G=n\, 2\pi/a$ -- this proportionality is strict in the absence of dipolar interactions. Since the stray field $B_{\rm sc}(x)$ scales linearly with $B_0$, then all its Fourier coefficients will scale at the same rate. This explains the almost linear dependence of the gap width on $B_0$ -- see also SI~2.

The dependence of the SW spectrum on the width of the gaps between the SC strips $d$, shown in Fig.~\ref{fig:B0_sep}(b), has unique features not observed for convectional magnonic crystals. Let's first discuss the effects that are not surprising, and are also present in many realizations of 1D magnonic crystals, that have a form of SW conduit (layer) containing a periodic sequence of centers (wells) that can partially confine the SWs.  In the limit of large separation between centers $d\rightarrow\infty$, we should obtain: (i) a quasi-continuum band spectrum for the frequencies larger than the FMR frequency of the pristine conduit $f>f_{\rm FMR}$, where both the gap widths and the band widths tend to zero, and (ii) a discrete spectrum of SW modes confined in the centers for $f<f_{\rm FMR}$. In the limit $d\rightarrow 0$, one can expect that the centers (in our case the wells of the stray field) will merge to form the uniform channel, and this will give a continuous SW spectrum with the closed gaps due to the absence of scattering. However, the peculiarity of our system is that the stray field generated by the sequence of SC strips separated by tiny gaps will still remain non-uniform, mainly due to the formation of narrow but relatively high barriers separating the wells. This effect will prevent the frequency gaps from closing for very small values of $d$. To illustrate this effect, we have plotted, in the SI~2, the changes of the widths of successive frequency gaps. It is clear that, unlike the $d\rightarrow\infty$ case where the frequency gaps gradually close, the gaps remain open for very small values of $d$. It is worth noting that the sweep of the bulk parameter $d$ in its entire range results in the intersection of the band edges and the closing of the gaps at certain values of $d$ -- see also SI~2.  

\section{Discussion}

We have studied the on-demand induction of MC in a uniform magnetic layer by placing a periodic sequence of SC strips above an FM layer. The MC can be activated or deactivated by changing the temperature below or above the critical temperature of the SC material in the presence uniform external field. 

Applying an out-of-plane external magnetic field to this superconductor--ferrimagnet hybrid structure induces a periodic field profile within the magnetic layer due to eddy currents generated in the SC strips by the Meissner effect. The depth of this periodic field profile varies linearly with the applied field, allowing the transition from weakly to strongly modulated MC and the control of the widths of the bands and gaps in the SW spectrum. Such control is also possible when temperature approaching the critical value and the periodic profile of the stray field gradually disappears due to the increase of the London penetration depth.

Furthermore, it has been shown that varying the distance between the SC strips and changing their width affects the stray-field profile in a complex and unintuitive manner. We have shown that adjusting the geometry of the SC pattern provides an additional degree of freedom to tailor the SW spectrum in the superconductor--ferrimagnet hybrid system, which is unprecedented in conventional magnonic crystals. In particular, we demonstrated that even for very narrow  gaps between adjacent SC strips the SW band gaps remained opened.

To describe the SW dynamics in this system, we have developed a theoretical approach to solve the London equation and the LL equation. Our theoretical calculations were validated by numerical simulations, and the results were in good agreement. We demonstrated that for an out-of-plane applied field, the SW spectrum in the superconductor--ferrimagnet hybrid is mostly determined by the static stray field generated by the SC pattern, and the effect of dynamical coupling can be neglected for this geometry.

The presented research demonstrates the possibility of inducing the magnonic structures on demand by the inhomogeneous magnetic field of a superconductor, which can be controlled by temperature or by a homogeneous external magnetic field. Systems of this type can be used as a reconfigurable platform for the SW processing.

\section{Methods}

\subsection*{The static magnetic field produced by the sequence of SC strips}

For the superconductor in Meissner state, the connection between magnetic vector potential $\mathbf{A}_{\rm sc}$ and SC current density $\mathbf{J}$, which appears as a response to the external magnetic field, is expressed in the London equation:
\begin{equation}
\centering
    \mathbf{A}_{\rm sc}=-\mu_0\lambda^2 \mathbf{J},\label{eq:london}
\end{equation}
where $\lambda$ is London penetration depth and $\mu_0$ is vacuum permeability.
The semi-analytical solution of the London equation for the rectangular strip was developed by Brandt \cite{Brandt1994a,Brandt1994b} and used to describe the hybrid structure of a SC strip over a FM film \cite{Kharlan2024}. For the SC strip of infinite length along the $z$-direction and placed in an external field applied along the $y$-direction $\mathbf{B}_0=B_0\hat{\mathbf{y}}$, the Meissner effect induces a current of density $\mathbf{J}=J(x,y)\hat{\mathbf{z}}$. The Amp\`ere law: $\Delta A_{\rm sc}=-\mu_0\mathbf{J}$ allows us to obtain the expression relating the magnetic vector potential and the current density (see \cite{Kharlan2024} for details of the derivation):

\begin{equation}
    \centering
        A_{\rm sc}(x,y)\!=-\mu_0\!\iint_{S}\! dx'dy' Q(x,y,x',y')
        J(x',y')+\!A_{\rm 0}(x),\label{eq:curr_int} 
\end{equation}   

where $S$ is the $xy$-crosssection of the strip, $A_{\rm sc}$ and $A_{\rm 0}$ are the magnetic vector potential contributions related to the magnetic field created by SC strip and an external magnetic field, respectively. The function $Q(x,y,x',y')=\tfrac{1}{2\pi}\ln\big(\sqrt{(x-x')^2+(y-y')^2}\big)$ is the integral kernel for two-dimensional Laplace operator. Using Eq.~(\ref{eq:london}) and setting $A_{\rm 0}=-B_0 x$ for the case of the single SC strip, Eq.~(\ref{eq:curr_int}) leads us to the integral equation for the current density $\lambda^2 J(x,y)\!=\!\!\!\iint_{S} dx'dy' Q(x,y,x',y')J(x',y')+\!x H_0$.

Equation~\ref{eq:curr_int} can be generalized for the case of a sequence of SC strips (not necessarily periodic) with an arbitrary amount of strips. Such transition should be done carefully by the correct choosing of the vector potential gauge. For a sufficiently large number of strips, as the system approaches infinity, both $A_{\rm sc}$ and $J$ become spatially periodic. At first sight, the linear variation of $A_{\rm sc}$ in space can lead to the misleading conclusion that the total magnetic vector potential and thus the current density does not have translation symmetry and depends on the coordinate origin. This contradiction disappears if we recall that $A$ is defined up to the gauge transformation \(A_{\rm sc} \rightarrow A_{\rm sc}+\nabla \mathbf{\phi}\) \cite{Tinkham_2004}, i.e., is not defined uniquely. However, the experimentally measurable current should be determined uniquely.

Let us consider the strip whose center is shifted along the $x$-axis by the distance $D$ with respect to the coordinate system origin. Taking $\nabla \mathbf{\phi} = D B_{0}$ for $A_{\rm sc}$, we can obtain the current density which is the same as for the strip located at the coordinate origin. 
Therefore, for the sequence of $2M+1$ strips with the period $a=w+d$ (see Fig.~\ref{fig:structure}), we should make transformation for $A_{\rm sc}$ for the $n$th strip ($n=-M,\ldots,M$) as $A_{{\rm sc},n} \rightarrow A_{{\rm sc},n}+B_0 a n$ \cite{Milosevic_2004,Doria_1989}. In this case, Eq.~(\ref{eq:curr_int}) gives us the formula for the current density:   
\begin{equation}
\centering
    \begin{split}
    \lambda^2 J(x,y)=\!\iint_{\Tilde{S}} dx'dy'Q(x,y,x',y')J(x',y')\\
    +(x- a n)\tfrac{1}{\mu_0} B_0,\label{eq:curr_int2}
    \end{split}     
\end{equation}
where integration should be taken over cross-section of all strips $\Tilde{S}$. Eq.~(\ref{eq:curr_int2}) can be integrated numerically by sampling the function $J(x, y)$ on a square grid of equidistant points (for details see \cite{Kharlan2024}).

Finally, the distribution of the field $B_{\rm sc}$ generated by the eddy currents of the sequence of SC strips can be determined from the Biot--Savart law:
\begin{equation}
\centering
    \begin{split}
    B_{{\rm sc},x}(x,y)=\frac{\mu_0}{2\pi}\iint_{\Tilde{S}} dx'dy'\frac{J(x',y')(y'-y)}{(x'-x)^2+(y'-y)^2},\\
    B_{{\rm sc},y}(x,y)=-\frac{\mu_0}{2\pi}\iint_{\Tilde{S}} dx'dy'\frac{J(x',y')(x'-x)}{(x'-x)^2+(y'-y)^2},\label{eq:b_field}
    \end{split}  
\end{equation}
where integration is performed over the cross-section $\Tilde{S}$ of all SC strips in the sequence.

The only material parameter for the London equation (Eq.~\ref{eq:london}) is the penetration depth $\lambda$ which, in general, varies with temperature. The model used does not predict transitions to a mixed and then normal state. Therefore, for our considerations, we are operating at the liquid-helium temperatures $0<T<T_{c}/2$. According to the textbook relation which is based on the two-fluid model \cite{Tinkham_2004}, we can estimate that the change in penetration depth $\lambda (T)$ is negligible if $T<T_{c}/2$. For the considered thickness of Nb strips (100~nm), the critical temperature takes the value characteristic for the bulk sample $T_{c,{\rm Nb}}=9.2~K$ \cite{Ilin_2020}.

Assumed low temperature ($T < T_{c}/2$) and the low external magnetic field prevents the system from transitioning to a mixed state. What is more, it is known that the external magnetic field needed for the transition to the mixed state depends on the width of the strip \cite{Kharlan2024}. Therefore, in our study, we chose to use low external magnetic fields ($25$~mT) for wide strips ($w=800$~nm) so as not to cause a transition to the mixed state. Without this approach, the potential emergence of vortices would diminish the strength of the stray field, thereby modifying the effects presented in this paper. Moreover, the considered system is periodic in one dimension, whereas the introduction of vortices would introduce periodicity along strips, complicating the interpretation of the results.

\subsection*{The SW mode propagation in periodic magnetic field induced by the sequence of SC strips}

To study the SW propagation in the FM film placed in the periodic stray field of the sequence of SC strips, we have used the PWM \cite{Krawczyk_2012}. This method is useful for calculating the excitation spectra in linear systems with discrete translational symmetry. In our theoretical model, we have made two assumptions: (i) the stray field produced by the sequence of SC strips does not depend on the $y$-coordinate and is equal to the value at the film center, which is reasonable for thin films placed in the close vicinity of the SC pattern; (ii) the tangential component of the SC field, which deviates the magnetization from the film normal, almost does not influence the SW frequencies and can be neglected \cite{Kharlan2024}. Therefore, we have included only the normal component of the SC field and considered the uniform magnetization directed perpendicular to the film plane.

The magnetization dynamics is described by LL equation, which takes the following form in the absence of damping:
\begin{equation}
\centering
    \frac{\partial \mathbf{M}(\mathbf{r},t)}{\partial t}=-|\gamma|\mu_0 \mathbf{M}(\mathbf{r},t)\times\mathbf{H}_{\rm eff}(\mathbf{r},t). \label{eq:LL}
\end{equation}
The symbol $\gamma$ denotes gyromagnetic ratio and $\mathbf{H}_{\rm eff}$ is effective magnetic field, calculated from the free energy density \cite{Gurevich1996}.
We considered the linear dynamics where the magnetization precession can be described as follows: $y$-component of the magnetization is constant and equal to the saturation magnetization $M_{\rm s}$, while a small, time-dependent component $\mathbf{m}=m_{x}\hat{\mathbf{x}}+m_{z}\hat{\mathbf{z}}$ is rotating in the $xz$-plane: $\mathbf{M}(x,y,t)=M_s\hat{\mathbf{y}}+\mathbf{m}(x,y)e^{i\omega t}$. The effective magnetic field can be presented as:
\begin{equation}
\centering
\begin{split}
    \mathbf{H}_{\rm eff}(x,y,t)&=  \Big(h_{{\rm d},x}(x,y)e^{i\omega t}+\lambda_{\rm ex}^2\Delta m_x(x,y)e^{i\omega t}\Big)\hat{\mathbf{x}}+\\
    &+\Big(H_0-M_{\rm s}+H_{\rm A}+H_{\rm sc}(x,y)\Big)\hat{\mathbf{y}}+\\ &+\Big(h_{{\rm d},z}(x,y)e^{i\omega t}\lambda_{\rm ex}^2\Delta m_z(x,y)e^{i\omega t}\Big)\hat{\mathbf{z}},
\end{split}
\label{eq:Heff}
\end{equation}
where $H_0$ is the external magnetic field; $h_{{\rm d},\alpha}(x,y)$ and $-M_{\rm s}$ are the dynamic and static components of the demagnetizing field, respectively ($\alpha={x,z}$). The field $H_{\rm A}=2K_{\rm u}/(\mu_0M_{\rm s}) $ is the perpendicular anisotropy field, and $H_{\rm sc}(x)=\mu_0 B_{{\rm sc},y}(x)$ is the stray field produced by SC strips (Eq.~(\ref{eq:b_field})). The terms $\lambda_{\rm ex}^2\Delta m_{\alpha}(x,y)$, where $\lambda_{\rm ex}$ is the exchange length, are the corresponding components ($\alpha={x,z}$) of dynamic exchange field. The dynamic demagnetizing field $\mathbf{h}_{{\rm d}}=h_{{\rm d},x}\hat{\mathbf{x}}+h_{{\rm d},z}\hat{\mathbf{z}}$ is dependent on the spatial profiles of dynamic magnetization, and is expressed in the nonlocal relation, presented here in the general form:
\begin{equation}
\centering
    \mathbf{h}_{\rm d}(\mathbf{r})=-\nabla\int_{V}dv'\mathbf{m}(\mathbf{r}')\cdot\nabla'\frac{1}{4\pi|\mathbf{r}-\mathbf{r}'|}.\label{eq:dyn_demag}
\end{equation}
Substituting Eqs.~(\ref{eq:Heff}) and (\ref{eq:dyn_demag}) into Eq.~(\ref{eq:LL}) gives us the linearized system of LL equations:
\begin{equation}
\centering
    \begin{split}
    i\Omega m_{k,x}(x)=\frac{M_{\rm s}}{\Tilde{H}_0}\left(h_{{\rm d},z}(x)+\lambda_{\rm ex}^2\Delta m_{k,z}(x)\right)\\
    -m_{k,z}(x)\left(1+\frac{H_{\rm sc}(x)}{\Tilde{H}_0}\right),\\
    i\Omega m_{k,z}(x)=-\frac{M_{\rm s}}{\Tilde{H}_0}\left(h_{{\rm d},x}(x)+\lambda_{\rm ex}^2\Delta m_{k,x}(x)\right)\\
    +m_{k,x}(x)\left(1+\frac{H_{\rm sc}(x)}{\Tilde{H}_0}\right),
     \label{eq:LL system}
     \end{split}
\end{equation}
where $\Omega = \omega/(\left| \gamma\right|\mu_0\tilde{H}_0)$ and $\tilde{H}_0=H_0+H_A-M_{\rm s}$. We assumed that both dynamic magnetization and dynamic field do not change significantly across the FM layer and took their values from the center $y=s-(t+d)/2$. The set of linear integro-differential equations (\ref{eq:dyn_demag}) and (\ref{eq:LL system}) has a periodic term $H_{\rm sc}(x)=H_{\rm sc}(x+a)$, where $a=w+d$ is a lattice constant of MC -- see Fig.~\ref{fig:structure}. Therefore, the solutions are Bloch functions depending on the wave vector $k$: $m_{k,\alpha}(x)=u_{k,\alpha}(x)e^{i k x}$, where $u_{k,\alpha}(x)=u_{k,\alpha}(x+a)$. Eqs.~(\ref{eq:LL system}) with auxiliary Eq.~(\ref{eq:dyn_demag}) form the eigenproblem for the Bloch functions $m_{k,\alpha}$ with corresponding eigenfrequencies $\Omega$. The PWM is based on the Fourier transform of the differential eigenproblem Eq.~(\ref{eq:LL system}). The Fourier expansion of the Bloch functions $m_{k,\alpha}$ and the periodic stray field $H_{\rm sc}$ transforms Eq.~(\ref{eq:LL system}) into an algebraic eigenvalue problem, which is much easier to solve:
\begin{equation}
\centering
i\Omega
    \begin{aligned}
    \begin{bmatrix}
      \mathbf{\bar{m}}_{k,x}\\
    \mathbf{\bar{m}}_{k,z}
    \end{bmatrix}
    \end{aligned}
    \begin{aligned}
    =
    \end{aligned}
    \begin{bmatrix}
        0  & \mathbf{\bar{\bar{M}}}_{xz}\\
        \mathbf{\bar{\bar{M}}}_{zx} & 0\\
    \end{bmatrix}
     \begin{aligned}
    \begin{bmatrix}
      \mathbf{\bar{m}}_{k,x}\\
    \mathbf{\bar{m}}_{k,z}
    \end{bmatrix}
    \end{aligned}
     \label{eq:evprob}.
\end{equation}
The vectors $\mathbf{\bar{m}}_{k,\alpha}=[\ldots,m_{k,\alpha,G},\ldots]^{T}$ are composed of the Fourier coefficients $m_{k,\alpha,G}$ for the expansion of the periodic factors of Bloch functions $u_{k,\alpha}(x)$. The implementation of PWM for the considered case, together with the explicit form of the matrices $\mathbf{\bar{\bar{M}}}_{xz}$ and $\mathbf{\bar{\bar{M}}}_{zx}$, is presented in SI~1.

The PWM results were cross-checked with FEM computations performed in COMSOL Multiphysics. The FEM simulations were performed including both out-of-plane and in-plane components of the static stray field generated by the sequence of SC strips. We relaxed the magnetic configuration, which resulted in a small tilt of the magnetization vector in the regions below the edges of the SC strips. Then equation (\ref{eq:LL system}) was automatically linearized for the obtained equilibrium configuration and solved with a Gauss equation for magnetism to include the static and dynamic demagnetization effects, within the magnetostatic approximation. Assuming harmonic dynamics in time, we solve the eigenvalue problem for successive values of the wave number.

It is worth noting that in the temperatures of a few kelvins, for which we consider the Meissner state of Nb strips, are still above the range of extremely low temperatures (a few tens of mK) where the spin-wave damping in the YIG layer (typically deposited on GGG substrate) starts to grow. Therefore, we can neglect the damping term in LL equation (\ref{eq:LL}). On the other hand, the material parameters for YIG (e.g. $M_{\rm s}$) will not change significantly in the liquid-helium temperatures, because even the room temperature is considered to be low, comparing the Curie temperature $\sim$500 K for YIG.

\section*{Author contributions statement}

J.K and J.W.K. conceived the research. J.K and J.W.K. performed theoretical background for this research. K.S and G.C designed and performed the numerical simulations. All authors analyzed the data, prepared the manuscript, and participated in its proofreading.

\section*{Competing interests}

The authors declare no competing interests.
\begin{acknowledgements}
The work was supported by the grants of the National Science Center, Poland, Nos. UMO-2019/35/D/ST3/03729, UMO-2020/39/O/ST5/02110, UMO-2021/43/I/ST3/00550, and UMO-2021/41/N/ST3/04478. G.C. would like to thank for the support from the Polish National Agency for Academic Exchange -- grant no. BPN/PRE/2022/1/00014/U/00001
\end{acknowledgements}

\section*{Data availability}
The data for the essential figures (Fig.~\ref{fig:SC field}, Fig.~\ref{fig:Dispersion_relations}, Fig.~\ref{fig:B0_sep}) can be accessed via the following \href{https://zenodo.org/doi/10.5281/zenodo.12742310}{URL}. The remaining data is available from the corresponding author on reasonable request.


\begin{thebibliography}{46}%
\makeatletter
\providecommand \@ifxundefined [1]{%
 \@ifx{#1\undefined}
}%
\providecommand \@ifnum [1]{%
 \ifnum #1\expandafter \@firstoftwo
 \else \expandafter \@secondoftwo
 \fi
}%
\providecommand \@ifx [1]{%
 \ifx #1\expandafter \@firstoftwo
 \else \expandafter \@secondoftwo
 \fi
}%
\providecommand \natexlab [1]{#1}%
\providecommand \enquote  [1]{``#1''}%
\providecommand \bibnamefont  [1]{#1}%
\providecommand \bibfnamefont [1]{#1}%
\providecommand \citenamefont [1]{#1}%
\providecommand \href@noop [0]{\@secondoftwo}%
\providecommand \href [0]{\begingroup \@sanitize@url \@href}%
\providecommand \@href[1]{\@@startlink{#1}\@@href}%
\providecommand \@@href[1]{\endgroup#1\@@endlink}%
\providecommand \@sanitize@url [0]{\catcode `\\12\catcode `\$12\catcode `\&12\catcode `\#12\catcode `\^12\catcode `\_12\catcode `\%12\relax}%
\providecommand \@@startlink[1]{}%
\providecommand \@@endlink[0]{}%
\providecommand \url  [0]{\begingroup\@sanitize@url \@url }%
\providecommand \@url [1]{\endgroup\@href {#1}{\urlprefix }}%
\providecommand \urlprefix  [0]{URL }%
\providecommand \Eprint [0]{\href }%
\providecommand \doibase [0]{https://doi.org/}%
\providecommand \selectlanguage [0]{\@gobble}%
\providecommand \bibinfo  [0]{\@secondoftwo}%
\providecommand \bibfield  [0]{\@secondoftwo}%
\providecommand \translation [1]{[#1]}%
\providecommand \BibitemOpen [0]{}%
\providecommand \bibitemStop [0]{}%
\providecommand \bibitemNoStop [0]{.\EOS\space}%
\providecommand \EOS [0]{\spacefactor3000\relax}%
\providecommand \BibitemShut  [1]{\csname bibitem#1\endcsname}%
\let\auto@bib@innerbib\@empty
\bibitem [{\citenamefont {Krawczyk}\ and\ \citenamefont {Grundler}(2014)}]{Krawczyk_2014}%
  \BibitemOpen
  \bibfield  {author} {\bibinfo {author} {\bibfnamefont {M.}~\bibnamefont {Krawczyk}}\ and\ \bibinfo {author} {\bibfnamefont {D.}~\bibnamefont {Grundler}},\ }\bibfield  {title} {\bibinfo {title} {Review and prospects of magnonic crystals and devices with reprogrammable band structure},\ }\href {https://doi.org/10.1088/0953-8984/26/12/123202} {\bibfield  {journal} {\bibinfo  {journal} {J. Phys. Condens. Matter}\ }\textbf {\bibinfo {volume} {26}},\ \bibinfo {pages} {123202} (\bibinfo {year} {2014})}\BibitemShut {NoStop}%
\bibitem [{\citenamefont {Kłos}\ \emph {et~al.}(2012)\citenamefont {Kłos}, \citenamefont {Sokolovskyy}, \citenamefont {Mamica},\ and\ \citenamefont {Krawczyk}}]{Klos_2012}%
  \BibitemOpen
  \bibfield  {author} {\bibinfo {author} {\bibfnamefont {J.~W.}\ \bibnamefont {Kłos}}, \bibinfo {author} {\bibfnamefont {M.~L.}\ \bibnamefont {Sokolovskyy}}, \bibinfo {author} {\bibfnamefont {S.}~\bibnamefont {Mamica}},\ and\ \bibinfo {author} {\bibfnamefont {M.}~\bibnamefont {Krawczyk}},\ }\bibfield  {title} {\bibinfo {title} {{The impact of the lattice symmetry and the inclusion shape on the spectrum of 2D magnonic crystals}},\ }\href {https://doi.org/10.1063/1.4729559} {\bibfield  {journal} {\bibinfo  {journal} {J. Appl. Phys.}\ }\textbf {\bibinfo {volume} {111}},\ \bibinfo {pages} {123910} (\bibinfo {year} {2012})}\BibitemShut {NoStop}%
\bibitem [{\citenamefont {Langer}\ \emph {et~al.}(2019)\citenamefont {Langer}, \citenamefont {Gallardo}, \citenamefont {Schneider}, \citenamefont {Stienen}, \citenamefont {Rold\'an-Molina}, \citenamefont {Yuan}, \citenamefont {Lenz}, \citenamefont {Lindner}, \citenamefont {Landeros},\ and\ \citenamefont {Fassbender}}]{Langer_2019}%
  \BibitemOpen
  \bibfield  {author} {\bibinfo {author} {\bibfnamefont {M.}~\bibnamefont {Langer}}, \bibinfo {author} {\bibfnamefont {R.~A.}\ \bibnamefont {Gallardo}}, \bibinfo {author} {\bibfnamefont {T.}~\bibnamefont {Schneider}}, \bibinfo {author} {\bibfnamefont {S.}~\bibnamefont {Stienen}}, \bibinfo {author} {\bibfnamefont {A.}~\bibnamefont {Rold\'an-Molina}}, \bibinfo {author} {\bibfnamefont {Y.}~\bibnamefont {Yuan}}, \bibinfo {author} {\bibfnamefont {K.}~\bibnamefont {Lenz}}, \bibinfo {author} {\bibfnamefont {J.}~\bibnamefont {Lindner}}, \bibinfo {author} {\bibfnamefont {P.}~\bibnamefont {Landeros}},\ and\ \bibinfo {author} {\bibfnamefont {J.}~\bibnamefont {Fassbender}},\ }\bibfield  {title} {\bibinfo {title} {Spin-wave modes in transition from a thin film to a full magnonic crystal},\ }\href {https://doi.org/10.1103/PhysRevB.99.024426} {\bibfield  {journal} {\bibinfo  {journal} {Phys. Rev. B}\ }\textbf {\bibinfo {volume} {99}},\ \bibinfo {pages} {024426} (\bibinfo {year} {2019})}\BibitemShut {NoStop}%
\bibitem [{\citenamefont {Choudhury}\ \emph {et~al.}(2020)\citenamefont {Choudhury}, \citenamefont {Chaurasiya}, \citenamefont {Mondal}, \citenamefont {Rana}, \citenamefont {Miura}, \citenamefont {Takahashi}, \citenamefont {Otani},\ and\ \citenamefont {Barman}}]{Choudhury_2020}%
  \BibitemOpen
  \bibfield  {author} {\bibinfo {author} {\bibfnamefont {S.}~\bibnamefont {Choudhury}}, \bibinfo {author} {\bibfnamefont {A.~K.}\ \bibnamefont {Chaurasiya}}, \bibinfo {author} {\bibfnamefont {A.~K.}\ \bibnamefont {Mondal}}, \bibinfo {author} {\bibfnamefont {B.}~\bibnamefont {Rana}}, \bibinfo {author} {\bibfnamefont {K.}~\bibnamefont {Miura}}, \bibinfo {author} {\bibfnamefont {H.}~\bibnamefont {Takahashi}}, \bibinfo {author} {\bibfnamefont {Y.}~\bibnamefont {Otani}},\ and\ \bibinfo {author} {\bibfnamefont {A.}~\bibnamefont {Barman}},\ }\bibfield  {title} {\bibinfo {title} {Voltage controlled on-demand magnonic nanochannels},\ }\href {https://doi.org/10.1126/sciadv.aba5457} {\bibfield  {journal} {\bibinfo  {journal} {Sci. Adv.}\ }\textbf {\bibinfo {volume} {6}},\ \bibinfo {pages} {eaba5457} (\bibinfo {year} {2020})}\BibitemShut {NoStop}%
\bibitem [{\citenamefont {Yu}\ \emph {et~al.}(2021)\citenamefont {Yu}, \citenamefont {Xiao},\ and\ \citenamefont {Schultheiss}}]{Yu_2021}%
  \BibitemOpen
  \bibfield  {author} {\bibinfo {author} {\bibfnamefont {H.}~\bibnamefont {Yu}}, \bibinfo {author} {\bibfnamefont {J.}~\bibnamefont {Xiao}},\ and\ \bibinfo {author} {\bibfnamefont {H.}~\bibnamefont {Schultheiss}},\ }\bibfield  {title} {\bibinfo {title} {Magnetic texture based magnonics},\ }\href {https://doi.org/https://doi.org/10.1016/j.physrep.2020.12.004} {\bibfield  {journal} {\bibinfo  {journal} {Phys. Rep.}\ }\textbf {\bibinfo {volume} {905}},\ \bibinfo {pages} {1} (\bibinfo {year} {2021})}\BibitemShut {NoStop}%
\bibitem [{\citenamefont {Szulc}\ \emph {et~al.}(2022)\citenamefont {Szulc}, \citenamefont {Tacchi}, \citenamefont {Hierro-Rodríguez}, \citenamefont {Díaz}, \citenamefont {Gruszecki}, \citenamefont {Graczyk}, \citenamefont {Quirós}, \citenamefont {Markó}, \citenamefont {Martín}, \citenamefont {Vélez}, \citenamefont {Schmool}, \citenamefont {Carlotti}, \citenamefont {Krawczyk},\ and\ \citenamefont {Álvarez Prado}}]{Szulc2022}%
  \BibitemOpen
  \bibfield  {author} {\bibinfo {author} {\bibfnamefont {K.}~\bibnamefont {Szulc}}, \bibinfo {author} {\bibfnamefont {S.}~\bibnamefont {Tacchi}}, \bibinfo {author} {\bibfnamefont {A.}~\bibnamefont {Hierro-Rodríguez}}, \bibinfo {author} {\bibfnamefont {J.}~\bibnamefont {Díaz}}, \bibinfo {author} {\bibfnamefont {P.}~\bibnamefont {Gruszecki}}, \bibinfo {author} {\bibfnamefont {P.}~\bibnamefont {Graczyk}}, \bibinfo {author} {\bibfnamefont {C.}~\bibnamefont {Quirós}}, \bibinfo {author} {\bibfnamefont {D.}~\bibnamefont {Markó}}, \bibinfo {author} {\bibfnamefont {J.~I.}\ \bibnamefont {Martín}}, \bibinfo {author} {\bibfnamefont {M.}~\bibnamefont {Vélez}}, \bibinfo {author} {\bibfnamefont {D.~S.}\ \bibnamefont {Schmool}}, \bibinfo {author} {\bibfnamefont {G.}~\bibnamefont {Carlotti}}, \bibinfo {author} {\bibfnamefont {M.}~\bibnamefont {Krawczyk}},\ and\ \bibinfo {author} {\bibfnamefont {L.~M.}\ \bibnamefont {Álvarez Prado}},\ }\bibfield  {title} {\bibinfo {title} {Reconfigurable magnonic crystals based on
  imprinted magnetization textures in hard and soft dipolar-coupled bilayers},\ }\href {https://doi.org/10.1021/acsnano.2c04256} {\bibfield  {journal} {\bibinfo  {journal} {ACS Nano}\ }\textbf {\bibinfo {volume} {16}},\ \bibinfo {pages} {14168} (\bibinfo {year} {2022})}\BibitemShut {NoStop}%
\bibitem [{\citenamefont {M\"uhlbauer}\ \emph {et~al.}(2009)\citenamefont {M\"uhlbauer}, \citenamefont {Binz}, \citenamefont {Jonietz}, \citenamefont {Pfleiderer}, \citenamefont {Rosch}, \citenamefont {Neubauer}, \citenamefont {Georgii},\ and\ \citenamefont {B\"oni}}]{Muhlbauer_2009}%
  \BibitemOpen
  \bibfield  {author} {\bibinfo {author} {\bibfnamefont {S.}~\bibnamefont {M\"uhlbauer}}, \bibinfo {author} {\bibfnamefont {B.}~\bibnamefont {Binz}}, \bibinfo {author} {\bibfnamefont {F.}~\bibnamefont {Jonietz}}, \bibinfo {author} {\bibfnamefont {C.}~\bibnamefont {Pfleiderer}}, \bibinfo {author} {\bibfnamefont {A.}~\bibnamefont {Rosch}}, \bibinfo {author} {\bibfnamefont {A.}~\bibnamefont {Neubauer}}, \bibinfo {author} {\bibfnamefont {R.}~\bibnamefont {Georgii}},\ and\ \bibinfo {author} {\bibfnamefont {P.}~\bibnamefont {B\"oni}},\ }\bibfield  {title} {\bibinfo {title} {Skyrmion lattice in a chiral magnet},\ }\href {https://doi.org/10.1126/science.1166767} {\bibfield  {journal} {\bibinfo  {journal} {Science}\ }\textbf {\bibinfo {volume} {323}},\ \bibinfo {pages} {915} (\bibinfo {year} {2009})}\BibitemShut {NoStop}%
\bibitem [{\citenamefont {Mruczkiewicz}\ \emph {et~al.}(2016)\citenamefont {Mruczkiewicz}, \citenamefont {Gruszecki}, \citenamefont {Zelent},\ and\ \citenamefont {Krawczyk}}]{Mruczkiewicz_2016}%
  \BibitemOpen
  \bibfield  {author} {\bibinfo {author} {\bibfnamefont {M.}~\bibnamefont {Mruczkiewicz}}, \bibinfo {author} {\bibfnamefont {P.}~\bibnamefont {Gruszecki}}, \bibinfo {author} {\bibfnamefont {M.}~\bibnamefont {Zelent}},\ and\ \bibinfo {author} {\bibfnamefont {M.}~\bibnamefont {Krawczyk}},\ }\bibfield  {title} {\bibinfo {title} {Collective dynamical skyrmion excitations in a magnonic crystal},\ }\href {https://doi.org/10.1103/PhysRevB.93.174429} {\bibfield  {journal} {\bibinfo  {journal} {Phys. Rev. B}\ }\textbf {\bibinfo {volume} {93}},\ \bibinfo {pages} {174429} (\bibinfo {year} {2016})}\BibitemShut {NoStop}%
\bibitem [{\citenamefont {Chumak}\ \emph {et~al.}(2017)\citenamefont {Chumak}, \citenamefont {Serga},\ and\ \citenamefont {Hillebrands}}]{Chumak_2017}%
  \BibitemOpen
  \bibfield  {author} {\bibinfo {author} {\bibfnamefont {A.~V.}\ \bibnamefont {Chumak}}, \bibinfo {author} {\bibfnamefont {A.~A.}\ \bibnamefont {Serga}},\ and\ \bibinfo {author} {\bibfnamefont {B.}~\bibnamefont {Hillebrands}},\ }\bibfield  {title} {\bibinfo {title} {Magnonic crystals for data processing},\ }\href {https://doi.org/10.1088/1361-6463/aa6a65} {\bibfield  {journal} {\bibinfo  {journal} {J. Phys. D: Appl. Phys.}\ }\textbf {\bibinfo {volume} {50}},\ \bibinfo {pages} {244001} (\bibinfo {year} {2017})}\BibitemShut {NoStop}%
\bibitem [{\citenamefont {Flebus}\ \emph {et~al.}(2024)\citenamefont {Flebus} \emph {et~al.}}]{Flebus_Roadmap_2024}%
  \BibitemOpen
  \bibfield  {author} {\bibinfo {author} {\bibfnamefont {B.}~\bibnamefont {Flebus}} \emph {et~al.},\ }\bibfield  {title} {\bibinfo {title} {The 2024 magnonics roadmap},\ }\href {https://doi.org/10.1088/1361-648X/ad399c} {\bibfield  {journal} {\bibinfo  {journal} {J. Phys. Condens. Matter}\ }\textbf {\bibinfo {volume} {36}},\ \bibinfo {pages} {363501} (\bibinfo {year} {2024})}\BibitemShut {NoStop}%
\bibitem [{\citenamefont {Topp}\ \emph {et~al.}(2010)\citenamefont {Topp}, \citenamefont {Heitmann}, \citenamefont {Kostylev},\ and\ \citenamefont {Grundler}}]{Grundler_2010}%
  \BibitemOpen
  \bibfield  {author} {\bibinfo {author} {\bibfnamefont {J.}~\bibnamefont {Topp}}, \bibinfo {author} {\bibfnamefont {D.}~\bibnamefont {Heitmann}}, \bibinfo {author} {\bibfnamefont {M.~P.}\ \bibnamefont {Kostylev}},\ and\ \bibinfo {author} {\bibfnamefont {D.}~\bibnamefont {Grundler}},\ }\bibfield  {title} {\bibinfo {title} {Making a reconfigurable artificial crystal by ordering bistable magnetic nanowires},\ }\href {https://doi.org/10.1103/PhysRevLett.104.207205} {\bibfield  {journal} {\bibinfo  {journal} {Phys. Rev. Lett.}\ }\textbf {\bibinfo {volume} {104}},\ \bibinfo {pages} {207205} (\bibinfo {year} {2010})}\BibitemShut {NoStop}%
\bibitem [{\citenamefont {Szulc}\ \emph {et~al.}(2019)\citenamefont {Szulc}, \citenamefont {Lisiecki}, \citenamefont {Makarov}, \citenamefont {Zelent}, \citenamefont {Ku\ifmmode~\acute{s}\else \'{s}\fi{}wik}, \citenamefont {G\l{}owi\ifmmode~\acute{n}\else \'{n}\fi{}ski}, \citenamefont {K\l{}os}, \citenamefont {M\"unzenberg}, \citenamefont {Gieniusz}, \citenamefont {Dubowik}, \citenamefont {Stobiecki},\ and\ \citenamefont {Krawczyk}}]{Szulc_2019}%
  \BibitemOpen
  \bibfield  {author} {\bibinfo {author} {\bibfnamefont {K.}~\bibnamefont {Szulc}}, \bibinfo {author} {\bibfnamefont {F.}~\bibnamefont {Lisiecki}}, \bibinfo {author} {\bibfnamefont {A.}~\bibnamefont {Makarov}}, \bibinfo {author} {\bibfnamefont {M.}~\bibnamefont {Zelent}}, \bibinfo {author} {\bibfnamefont {P.}~\bibnamefont {Ku\ifmmode~\acute{s}\else \'{s}\fi{}wik}}, \bibinfo {author} {\bibfnamefont {H.}~\bibnamefont {G\l{}owi\ifmmode~\acute{n}\else \'{n}\fi{}ski}}, \bibinfo {author} {\bibfnamefont {J.~W.}\ \bibnamefont {K\l{}os}}, \bibinfo {author} {\bibfnamefont {M.}~\bibnamefont {M\"unzenberg}}, \bibinfo {author} {\bibfnamefont {R.}~\bibnamefont {Gieniusz}}, \bibinfo {author} {\bibfnamefont {J.}~\bibnamefont {Dubowik}}, \bibinfo {author} {\bibfnamefont {F.}~\bibnamefont {Stobiecki}},\ and\ \bibinfo {author} {\bibfnamefont {M.}~\bibnamefont {Krawczyk}},\ }\bibfield  {title} {\bibinfo {title} {Remagnetization in arrays of ferromagnetic nanostripes with periodic and quasiperiodic order},\ }\href
  {https://doi.org/10.1103/PhysRevB.99.064412} {\bibfield  {journal} {\bibinfo  {journal} {Phys. Rev. B}\ }\textbf {\bibinfo {volume} {99}},\ \bibinfo {pages} {064412} (\bibinfo {year} {2019})}\BibitemShut {NoStop}%
\bibitem [{\citenamefont {Gartside}\ \emph {et~al.}(2021)\citenamefont {Gartside}, \citenamefont {Vanstone}, \citenamefont {Dion}, \citenamefont {Stenning}, \citenamefont {Arroo}, \citenamefont {Kurebayashi},\ and\ \citenamefont {Branford}}]{Gartside_2021}%
  \BibitemOpen
  \bibfield  {author} {\bibinfo {author} {\bibfnamefont {J.~C.}\ \bibnamefont {Gartside}}, \bibinfo {author} {\bibfnamefont {A.}~\bibnamefont {Vanstone}}, \bibinfo {author} {\bibfnamefont {T.}~\bibnamefont {Dion}}, \bibinfo {author} {\bibfnamefont {K.~D.}\ \bibnamefont {Stenning}}, \bibinfo {author} {\bibfnamefont {D.~M.}\ \bibnamefont {Arroo}}, \bibinfo {author} {\bibfnamefont {H.}~\bibnamefont {Kurebayashi}},\ and\ \bibinfo {author} {\bibfnamefont {W.~R.}\ \bibnamefont {Branford}},\ }\bibfield  {title} {\bibinfo {title} {Reconfigurable magnonic mode-hybridisation and spectral control in a bicomponent artificial spin ice},\ }\href {https://doi.org/10.1038/s41467-021-22723-x} {\bibfield  {journal} {\bibinfo  {journal} {Nat. Commun.}\ }\textbf {\bibinfo {volume} {12}},\ \bibinfo {pages} {2488} (\bibinfo {year} {2021})}\BibitemShut {NoStop}%
\bibitem [{\citenamefont {Banerjee}\ \emph {et~al.}(2017)\citenamefont {Banerjee}, \citenamefont {Gruszecki}, \citenamefont {Klos}, \citenamefont {Hellwig}, \citenamefont {Krawczyk},\ and\ \citenamefont {Barman}}]{Gruszecki2017}%
  \BibitemOpen
  \bibfield  {author} {\bibinfo {author} {\bibfnamefont {C.}~\bibnamefont {Banerjee}}, \bibinfo {author} {\bibfnamefont {P.}~\bibnamefont {Gruszecki}}, \bibinfo {author} {\bibfnamefont {J.~W.}\ \bibnamefont {Klos}}, \bibinfo {author} {\bibfnamefont {O.}~\bibnamefont {Hellwig}}, \bibinfo {author} {\bibfnamefont {M.}~\bibnamefont {Krawczyk}},\ and\ \bibinfo {author} {\bibfnamefont {A.}~\bibnamefont {Barman}},\ }\bibfield  {title} {\bibinfo {title} {{Magnonic band structure in a Co/Pd stripe domain system investigated by Brillouin light scattering and micromagnetic simulations}},\ }\href {https://doi.org/10.1103/PhysRevB.96.024421} {\bibfield  {journal} {\bibinfo  {journal} {Phys. Rev. B}\ }\textbf {\bibinfo {volume} {96}},\ \bibinfo {pages} {024421} (\bibinfo {year} {2017})}\BibitemShut {NoStop}%
\bibitem [{\citenamefont {Wang}\ \emph {et~al.}(2017)\citenamefont {Wang}, \citenamefont {Chumak}, \citenamefont {Jin}, \citenamefont {Zhang}, \citenamefont {Hillebrands},\ and\ \citenamefont {Zhong}}]{Wang_2017}%
  \BibitemOpen
  \bibfield  {author} {\bibinfo {author} {\bibfnamefont {Q.}~\bibnamefont {Wang}}, \bibinfo {author} {\bibfnamefont {A.~V.}\ \bibnamefont {Chumak}}, \bibinfo {author} {\bibfnamefont {L.}~\bibnamefont {Jin}}, \bibinfo {author} {\bibfnamefont {H.}~\bibnamefont {Zhang}}, \bibinfo {author} {\bibfnamefont {B.}~\bibnamefont {Hillebrands}},\ and\ \bibinfo {author} {\bibfnamefont {Z.}~\bibnamefont {Zhong}},\ }\bibfield  {title} {\bibinfo {title} {Voltage-controlled nanoscale reconfigurable magnonic crystal},\ }\href {https://doi.org/10.1103/PhysRevB.95.134433} {\bibfield  {journal} {\bibinfo  {journal} {Phys. Rev. B}\ }\textbf {\bibinfo {volume} {95}},\ \bibinfo {pages} {134433} (\bibinfo {year} {2017})}\BibitemShut {NoStop}%
\bibitem [{\citenamefont {Rana}\ and\ \citenamefont {Otani}(2018)}]{Rana_2018}%
  \BibitemOpen
  \bibfield  {author} {\bibinfo {author} {\bibfnamefont {B.}~\bibnamefont {Rana}}\ and\ \bibinfo {author} {\bibfnamefont {Y.}~\bibnamefont {Otani}},\ }\bibfield  {title} {\bibinfo {title} {Voltage-controlled reconfigurable spin-wave nanochannels and logic devices},\ }\href {https://doi.org/10.1103/PhysRevApplied.9.014033} {\bibfield  {journal} {\bibinfo  {journal} {Phys. Rev. Appl.}\ }\textbf {\bibinfo {volume} {9}},\ \bibinfo {pages} {014033} (\bibinfo {year} {2018})}\BibitemShut {NoStop}%
\bibitem [{\citenamefont {Dahir}\ \emph {et~al.}(2020)\citenamefont {Dahir}, \citenamefont {Volkov},\ and\ \citenamefont {Eremin}}]{Dahir_2020}%
  \BibitemOpen
  \bibfield  {author} {\bibinfo {author} {\bibfnamefont {S.~M.}\ \bibnamefont {Dahir}}, \bibinfo {author} {\bibfnamefont {A.~F.}\ \bibnamefont {Volkov}},\ and\ \bibinfo {author} {\bibfnamefont {I.~M.}\ \bibnamefont {Eremin}},\ }\bibfield  {title} {\bibinfo {title} {Meissner currents induced by topological magnetic textures in hybrid superconductor/ferromagnet structures},\ }\href {https://doi.org/10.1103/PhysRevB.102.014503} {\bibfield  {journal} {\bibinfo  {journal} {Phys. Rev. B}\ }\textbf {\bibinfo {volume} {102}},\ \bibinfo {pages} {014503} (\bibinfo {year} {2020})}\BibitemShut {NoStop}%
\bibitem [{\citenamefont {Gonz\'alez-G\'omez}\ \emph {et~al.}(2022)\citenamefont {Gonz\'alez-G\'omez}, \citenamefont {Castell-Queralt}, \citenamefont {Del-Valle},\ and\ \citenamefont {Navau}}]{Gonzalez_2022}%
  \BibitemOpen
  \bibfield  {author} {\bibinfo {author} {\bibfnamefont {L.}~\bibnamefont {Gonz\'alez-G\'omez}}, \bibinfo {author} {\bibfnamefont {J.}~\bibnamefont {Castell-Queralt}}, \bibinfo {author} {\bibfnamefont {N.}~\bibnamefont {Del-Valle}},\ and\ \bibinfo {author} {\bibfnamefont {C.}~\bibnamefont {Navau}},\ }\bibfield  {title} {\bibinfo {title} {Mutual interaction between superconductors and ferromagnetic skyrmionic structures in confined geometries},\ }\href {https://doi.org/10.1103/PhysRevApplied.17.034069} {\bibfield  {journal} {\bibinfo  {journal} {Phys. Rev. Appl.}\ }\textbf {\bibinfo {volume} {17}},\ \bibinfo {pages} {034069} (\bibinfo {year} {2022})}\BibitemShut {NoStop}%
\bibitem [{\citenamefont {Palau}\ \emph {et~al.}(2016)\citenamefont {Palau}, \citenamefont {Valencia}, \citenamefont {Del-Valle}, \citenamefont {Navau}, \citenamefont {Cialone}, \citenamefont {Arora}, \citenamefont {Kronast}, \citenamefont {Tennant}, \citenamefont {Obradors}, \citenamefont {Sanchez},\ and\ \citenamefont {Puig}}]{Palau_2016}%
  \BibitemOpen
  \bibfield  {author} {\bibinfo {author} {\bibfnamefont {A.}~\bibnamefont {Palau}}, \bibinfo {author} {\bibfnamefont {S.}~\bibnamefont {Valencia}}, \bibinfo {author} {\bibfnamefont {N.}~\bibnamefont {Del-Valle}}, \bibinfo {author} {\bibfnamefont {C.}~\bibnamefont {Navau}}, \bibinfo {author} {\bibfnamefont {M.}~\bibnamefont {Cialone}}, \bibinfo {author} {\bibfnamefont {A.}~\bibnamefont {Arora}}, \bibinfo {author} {\bibfnamefont {F.}~\bibnamefont {Kronast}}, \bibinfo {author} {\bibfnamefont {D.~A.}\ \bibnamefont {Tennant}}, \bibinfo {author} {\bibfnamefont {X.}~\bibnamefont {Obradors}}, \bibinfo {author} {\bibfnamefont {A.}~\bibnamefont {Sanchez}},\ and\ \bibinfo {author} {\bibfnamefont {T.}~\bibnamefont {Puig}},\ }\bibfield  {title} {\bibinfo {title} {Encoding magnetic states in monopole-like configurations using superconducting dots},\ }\href {https://doi.org/https://doi.org/10.1002/advs.201600207} {\bibfield  {journal} {\bibinfo  {journal} {Adv. Sci.}\ }\textbf {\bibinfo {volume} {3}},\ \bibinfo {pages}
  {1600207} (\bibinfo {year} {2016})}\BibitemShut {NoStop}%
\bibitem [{\citenamefont {Golovchanskiy}\ \emph {et~al.}(2018)\citenamefont {Golovchanskiy}, \citenamefont {Abramov}, \citenamefont {Stolyarov}, \citenamefont {Bolginov}, \citenamefont {Ryazanov}, \citenamefont {Golubov},\ and\ \citenamefont {Ustinov}}]{Golovchanskiy2018}%
  \BibitemOpen
  \bibfield  {author} {\bibinfo {author} {\bibfnamefont {I.~A.}\ \bibnamefont {Golovchanskiy}}, \bibinfo {author} {\bibfnamefont {N.~N.}\ \bibnamefont {Abramov}}, \bibinfo {author} {\bibfnamefont {V.~S.}\ \bibnamefont {Stolyarov}}, \bibinfo {author} {\bibfnamefont {V.~V.}\ \bibnamefont {Bolginov}}, \bibinfo {author} {\bibfnamefont {V.~V.}\ \bibnamefont {Ryazanov}}, \bibinfo {author} {\bibfnamefont {A.~A.}\ \bibnamefont {Golubov}},\ and\ \bibinfo {author} {\bibfnamefont {A.~V.}\ \bibnamefont {Ustinov}},\ }\bibfield  {title} {\bibinfo {title} {Ferromagnet/superconductor hybridization for magnonic applications},\ }\href {https://doi.org/https://doi.org/10.1002/adfm.201802375} {\bibfield  {journal} {\bibinfo  {journal} {Adv. Funct. Mater.}\ }\textbf {\bibinfo {volume} {28}},\ \bibinfo {pages} {1802375} (\bibinfo {year} {2018})}\BibitemShut {NoStop}%
\bibitem [{\citenamefont {Borst}\ \emph {et~al.}(2023)\citenamefont {Borst}, \citenamefont {Vree}, \citenamefont {Lowther}, \citenamefont {Teepe}, \citenamefont {Kurdi}, \citenamefont {Bertelli}, \citenamefont {Simon}, \citenamefont {Blanter},\ and\ \citenamefont {Van Der~Sar}}]{borst_2023}%
  \BibitemOpen
  \bibfield  {author} {\bibinfo {author} {\bibfnamefont {M.}~\bibnamefont {Borst}}, \bibinfo {author} {\bibfnamefont {P.~H.}\ \bibnamefont {Vree}}, \bibinfo {author} {\bibfnamefont {A.}~\bibnamefont {Lowther}}, \bibinfo {author} {\bibfnamefont {A.}~\bibnamefont {Teepe}}, \bibinfo {author} {\bibfnamefont {S.}~\bibnamefont {Kurdi}}, \bibinfo {author} {\bibfnamefont {I.}~\bibnamefont {Bertelli}}, \bibinfo {author} {\bibfnamefont {B.~G.}\ \bibnamefont {Simon}}, \bibinfo {author} {\bibfnamefont {Y.~M.}\ \bibnamefont {Blanter}},\ and\ \bibinfo {author} {\bibfnamefont {T.}~\bibnamefont {Van Der~Sar}},\ }\bibfield  {title} {\bibinfo {title} {Observation and control of hybrid spin-wave–{Meissner}-current transport modes},\ }\href {https://doi.org/10.1126/science.adj7576} {\bibfield  {journal} {\bibinfo  {journal} {Science}\ }\textbf {\bibinfo {volume} {382}},\ \bibinfo {pages} {430} (\bibinfo {year} {2023})}\BibitemShut {NoStop}%
\bibitem [{\citenamefont {Kharlan}\ \emph {et~al.}(2024)\citenamefont {Kharlan}, \citenamefont {Sobucki}, \citenamefont {Szulc}, \citenamefont {Memarzadeh},\ and\ \citenamefont {Kłos}}]{Kharlan2024}%
  \BibitemOpen
  \bibfield  {author} {\bibinfo {author} {\bibfnamefont {J.}~\bibnamefont {Kharlan}}, \bibinfo {author} {\bibfnamefont {K.}~\bibnamefont {Sobucki}}, \bibinfo {author} {\bibfnamefont {K.}~\bibnamefont {Szulc}}, \bibinfo {author} {\bibfnamefont {S.}~\bibnamefont {Memarzadeh}},\ and\ \bibinfo {author} {\bibfnamefont {J.~W.}\ \bibnamefont {Kłos}},\ }\bibfield  {title} {\bibinfo {title} {Spin-wave confinement in a hybrid superconductor-ferrimagnet nanostructure},\ }\href {https://doi.org/10.1103/PhysRevApplied.21.064007} {\bibfield  {journal} {\bibinfo  {journal} {Phys. Rev. Appl.}\ }\textbf {\bibinfo {volume} {21}},\ \bibinfo {pages} {064007} (\bibinfo {year} {2024})}\BibitemShut {NoStop}%
\bibitem [{\citenamefont {Milo\ifmmode \check{s}\else \v{s}\fi{}evi\ifmmode~\acute{c}\else \'{c}\fi{}}\ and\ \citenamefont {Peeters}(2004)}]{Milosevic_2004}%
  \BibitemOpen
  \bibfield  {author} {\bibinfo {author} {\bibfnamefont {M.~V.}\ \bibnamefont {Milo\ifmmode \check{s}\else \v{s}\fi{}evi\ifmmode~\acute{c}\else \'{c}\fi{}}}\ and\ \bibinfo {author} {\bibfnamefont {F.~M.}\ \bibnamefont {Peeters}},\ }\bibfield  {title} {\bibinfo {title} {Vortex-antivortex lattices in superconducting films with magnetic pinning arrays},\ }\href {https://doi.org/10.1103/PhysRevLett.93.267006} {\bibfield  {journal} {\bibinfo  {journal} {Phys. Rev. Lett.}\ }\textbf {\bibinfo {volume} {93}},\ \bibinfo {pages} {267006} (\bibinfo {year} {2004})}\BibitemShut {NoStop}%
\bibitem [{\citenamefont {Khaydukov}\ \emph {et~al.}(2019)\citenamefont {Khaydukov}, \citenamefont {Kravtsov}, \citenamefont {Zhaketov}, \citenamefont {Progliado}, \citenamefont {Kim}, \citenamefont {Nikitenko}, \citenamefont {Keller}, \citenamefont {Ustinov}, \citenamefont {Aksenov},\ and\ \citenamefont {Keimer}}]{Khaydukov_2019}%
  \BibitemOpen
  \bibfield  {author} {\bibinfo {author} {\bibfnamefont {Y.~N.}\ \bibnamefont {Khaydukov}}, \bibinfo {author} {\bibfnamefont {E.~A.}\ \bibnamefont {Kravtsov}}, \bibinfo {author} {\bibfnamefont {V.~D.}\ \bibnamefont {Zhaketov}}, \bibinfo {author} {\bibfnamefont {V.~V.}\ \bibnamefont {Progliado}}, \bibinfo {author} {\bibfnamefont {G.}~\bibnamefont {Kim}}, \bibinfo {author} {\bibfnamefont {Y.~V.}\ \bibnamefont {Nikitenko}}, \bibinfo {author} {\bibfnamefont {T.}~\bibnamefont {Keller}}, \bibinfo {author} {\bibfnamefont {V.~V.}\ \bibnamefont {Ustinov}}, \bibinfo {author} {\bibfnamefont {V.~L.}\ \bibnamefont {Aksenov}},\ and\ \bibinfo {author} {\bibfnamefont {B.}~\bibnamefont {Keimer}},\ }\bibfield  {title} {\bibinfo {title} {Magnetic proximity effect in nb/gd superlattices seen by neutron reflectometry},\ }\href {https://doi.org/10.1103/PhysRevB.99.140503} {\bibfield  {journal} {\bibinfo  {journal} {Phys. Rev. B}\ }\textbf {\bibinfo {volume} {99}},\ \bibinfo {pages} {140503} (\bibinfo {year} {2019})}\BibitemShut
  {NoStop}%
\bibitem [{\citenamefont {Dobrovolskiy}\ \emph {et~al.}(2019)\citenamefont {Dobrovolskiy}, \citenamefont {Sachser}, \citenamefont {Brächer}, \citenamefont {Böttcher}, \citenamefont {Kruglyak}, \citenamefont {Vovk}, \citenamefont {Shklovskij}, \citenamefont {Huth}, \citenamefont {Hillebrands},\ and\ \citenamefont {Chumak}}]{Dobrovolskiy_2019}%
  \BibitemOpen
  \bibfield  {author} {\bibinfo {author} {\bibfnamefont {O.~V.}\ \bibnamefont {Dobrovolskiy}}, \bibinfo {author} {\bibfnamefont {R.}~\bibnamefont {Sachser}}, \bibinfo {author} {\bibfnamefont {T.}~\bibnamefont {Brächer}}, \bibinfo {author} {\bibfnamefont {T.}~\bibnamefont {Böttcher}}, \bibinfo {author} {\bibfnamefont {V.~V.}\ \bibnamefont {Kruglyak}}, \bibinfo {author} {\bibfnamefont {R.~V.}\ \bibnamefont {Vovk}}, \bibinfo {author} {\bibfnamefont {V.~A.}\ \bibnamefont {Shklovskij}}, \bibinfo {author} {\bibfnamefont {M.}~\bibnamefont {Huth}}, \bibinfo {author} {\bibfnamefont {B.}~\bibnamefont {Hillebrands}},\ and\ \bibinfo {author} {\bibfnamefont {A.~V.}\ \bibnamefont {Chumak}},\ }\bibfield  {title} {\bibinfo {title} {Magnon–fluxon interaction in a ferromagnet/superconductor heterostructure},\ }\href {https://doi.org/10.1038/s41567-019-0428-5} {\bibfield  {journal} {\bibinfo  {journal} {Nat. Phys.}\ }\textbf {\bibinfo {volume} {15}},\ \bibinfo {pages} {477} (\bibinfo {year} {2019})}\BibitemShut {NoStop}%
\bibitem [{\citenamefont {Jafri}\ \emph {et~al.}(2020)\citenamefont {Jafri}, \citenamefont {Huang}, \citenamefont {Yang}, \citenamefont {Wang}, \citenamefont {Amirov}, \citenamefont {Chen},\ and\ \citenamefont {Nan}}]{Jafri_2020}%
  \BibitemOpen
  \bibfield  {author} {\bibinfo {author} {\bibfnamefont {H.~M.}\ \bibnamefont {Jafri}}, \bibinfo {author} {\bibfnamefont {H.}~\bibnamefont {Huang}}, \bibinfo {author} {\bibfnamefont {C.}~\bibnamefont {Yang}}, \bibinfo {author} {\bibfnamefont {J.}~\bibnamefont {Wang}}, \bibinfo {author} {\bibfnamefont {A.~A.}\ \bibnamefont {Amirov}}, \bibinfo {author} {\bibfnamefont {L.-Q.}\ \bibnamefont {Chen}},\ and\ \bibinfo {author} {\bibfnamefont {C.-W.}\ \bibnamefont {Nan}},\ }\bibfield  {title} {\bibinfo {title} {Domain wall tuned superconductivity in superconductor–ferromagnet bilayers},\ }\href {https://doi.org/10.1088/1361-6463/ab932f} {\bibfield  {journal} {\bibinfo  {journal} {J. Phys. D: Appl. Phys.}\ }\textbf {\bibinfo {volume} {53}},\ \bibinfo {pages} {375001} (\bibinfo {year} {2020})}\BibitemShut {NoStop}%
\bibitem [{\citenamefont {Putilov}\ \emph {et~al.}(2022)\citenamefont {Putilov}, \citenamefont {Mironov}, \citenamefont {Mel'nikov},\ and\ \citenamefont {Buzdin}}]{Putilov_2022}%
  \BibitemOpen
  \bibfield  {author} {\bibinfo {author} {\bibfnamefont {A.~V.}\ \bibnamefont {Putilov}}, \bibinfo {author} {\bibfnamefont {S.~V.}\ \bibnamefont {Mironov}}, \bibinfo {author} {\bibfnamefont {A.~S.}\ \bibnamefont {Mel'nikov}},\ and\ \bibinfo {author} {\bibfnamefont {A.~I.}\ \bibnamefont {Buzdin}},\ }\bibfield  {title} {\bibinfo {title} {Giant electromagnetic proximity effect in superconductor/ferromagnet superlattices},\ }\href {https://doi.org/10.1103/PhysRevB.105.064510} {\bibfield  {journal} {\bibinfo  {journal} {Phys. Rev. B}\ }\textbf {\bibinfo {volume} {105}},\ \bibinfo {pages} {064510} (\bibinfo {year} {2022})}\BibitemShut {NoStop}%
\bibitem [{\citenamefont {Golovchanskiy}\ \emph {et~al.}(2019)\citenamefont {Golovchanskiy}, \citenamefont {Abramov}, \citenamefont {Stolyarov}, \citenamefont {Dzhumaev}, \citenamefont {Emelyanova}, \citenamefont {Golubov}, \citenamefont {Ryazanov},\ and\ \citenamefont {Ustinov}}]{Golovchanskiy2019}%
  \BibitemOpen
  \bibfield  {author} {\bibinfo {author} {\bibfnamefont {I.~A.}\ \bibnamefont {Golovchanskiy}}, \bibinfo {author} {\bibfnamefont {N.~N.}\ \bibnamefont {Abramov}}, \bibinfo {author} {\bibfnamefont {V.~S.}\ \bibnamefont {Stolyarov}}, \bibinfo {author} {\bibfnamefont {P.~S.}\ \bibnamefont {Dzhumaev}}, \bibinfo {author} {\bibfnamefont {O.~V.}\ \bibnamefont {Emelyanova}}, \bibinfo {author} {\bibfnamefont {A.~A.}\ \bibnamefont {Golubov}}, \bibinfo {author} {\bibfnamefont {V.~V.}\ \bibnamefont {Ryazanov}},\ and\ \bibinfo {author} {\bibfnamefont {A.~V.}\ \bibnamefont {Ustinov}},\ }\bibfield  {title} {\bibinfo {title} {Ferromagnet/superconductor hybrid magnonic metamaterials},\ }\href {https://doi.org/https://doi.org/10.1002/advs.201900435} {\bibfield  {journal} {\bibinfo  {journal} {Adv. Sci.}\ }\textbf {\bibinfo {volume} {6}},\ \bibinfo {pages} {1900435} (\bibinfo {year} {2019})}\BibitemShut {NoStop}%
\bibitem [{\citenamefont {Golovchanskiy}\ \emph {et~al.}(2020)\citenamefont {Golovchanskiy}, \citenamefont {Abramov}, \citenamefont {Stolyarov}, \citenamefont {Golubov}, \citenamefont {Ryazanov},\ and\ \citenamefont {Ustinov}}]{Golovchanskiy2020a}%
  \BibitemOpen
  \bibfield  {author} {\bibinfo {author} {\bibfnamefont {I.~A.}\ \bibnamefont {Golovchanskiy}}, \bibinfo {author} {\bibfnamefont {N.~N.}\ \bibnamefont {Abramov}}, \bibinfo {author} {\bibfnamefont {V.~S.}\ \bibnamefont {Stolyarov}}, \bibinfo {author} {\bibfnamefont {A.~A.}\ \bibnamefont {Golubov}}, \bibinfo {author} {\bibfnamefont {V.~V.}\ \bibnamefont {Ryazanov}},\ and\ \bibinfo {author} {\bibfnamefont {A.~V.}\ \bibnamefont {Ustinov}},\ }\bibfield  {title} {\bibinfo {title} {{Nonlinear spin waves in ferromagnetic/superconductor hybrids}},\ }\href {https://doi.org/10.1063/1.5141793} {\bibfield  {journal} {\bibinfo  {journal} {J. Appl. Phys.}\ }\textbf {\bibinfo {volume} {127}},\ \bibinfo {pages} {093903} (\bibinfo {year} {2020})}\BibitemShut {NoStop}%
\bibitem [{\citenamefont {Böttcher}\ \emph {et~al.}(2022)\citenamefont {Böttcher}, \citenamefont {Ruhwede}, \citenamefont {Levchenko}, \citenamefont {Wang}, \citenamefont {Chumak}, \citenamefont {Popov}, \citenamefont {Zavislyak}, \citenamefont {Dubs}, \citenamefont {Surzhenko}, \citenamefont {Hillebrands}, \citenamefont {Chumak},\ and\ \citenamefont {Pirro}}]{Bottcher2022}%
  \BibitemOpen
  \bibfield  {author} {\bibinfo {author} {\bibfnamefont {T.}~\bibnamefont {Böttcher}}, \bibinfo {author} {\bibfnamefont {M.}~\bibnamefont {Ruhwede}}, \bibinfo {author} {\bibfnamefont {K.}~\bibnamefont {Levchenko}}, \bibinfo {author} {\bibfnamefont {Q.}~\bibnamefont {Wang}}, \bibinfo {author} {\bibfnamefont {H.}~\bibnamefont {Chumak}}, \bibinfo {author} {\bibfnamefont {M.}~\bibnamefont {Popov}}, \bibinfo {author} {\bibfnamefont {I.}~\bibnamefont {Zavislyak}}, \bibinfo {author} {\bibfnamefont {C.}~\bibnamefont {Dubs}}, \bibinfo {author} {\bibfnamefont {O.}~\bibnamefont {Surzhenko}}, \bibinfo {author} {\bibfnamefont {B.}~\bibnamefont {Hillebrands}}, \bibinfo {author} {\bibfnamefont {A.}~\bibnamefont {Chumak}},\ and\ \bibinfo {author} {\bibfnamefont {P.}~\bibnamefont {Pirro}},\ }\bibfield  {title} {\bibinfo {title} {Fast long-wavelength exchange spin waves in partially compensated {Ga:YIG}},\ }\href {https://doi.org/10.1063/5.0082724} {\bibfield  {journal} {\bibinfo  {journal} {Appl. Phys. Lett.}\ }\textbf
  {\bibinfo {volume} {120}},\ \bibinfo {pages} {102401} (\bibinfo {year} {2022})}\BibitemShut {NoStop}%
\bibitem [{\citenamefont {Kharlan}\ \emph {et~al.}(2019)\citenamefont {Kharlan}, \citenamefont {Bondarenko}, \citenamefont {Krawczyk}, \citenamefont {Salyuk}, \citenamefont {Tartakovskaya}, \citenamefont {Trzaskowska},\ and\ \citenamefont {Golub}}]{Kharlan2019}%
  \BibitemOpen
  \bibfield  {author} {\bibinfo {author} {\bibfnamefont {J.}~\bibnamefont {Kharlan}}, \bibinfo {author} {\bibfnamefont {P.}~\bibnamefont {Bondarenko}}, \bibinfo {author} {\bibfnamefont {M.}~\bibnamefont {Krawczyk}}, \bibinfo {author} {\bibfnamefont {O.}~\bibnamefont {Salyuk}}, \bibinfo {author} {\bibfnamefont {E.}~\bibnamefont {Tartakovskaya}}, \bibinfo {author} {\bibfnamefont {A.}~\bibnamefont {Trzaskowska}},\ and\ \bibinfo {author} {\bibfnamefont {V.}~\bibnamefont {Golub}},\ }\bibfield  {title} {\bibinfo {title} {Standing spin waves in perpendicularly magnetized triangular dots},\ }\href {https://doi.org/10.1103/PhysRevB.100.184416} {\bibfield  {journal} {\bibinfo  {journal} {Phys. Rev. B}\ }\textbf {\bibinfo {volume} {100}},\ \bibinfo {pages} {184416} (\bibinfo {year} {2019})}\BibitemShut {NoStop}%
\bibitem [{\citenamefont {Gubin}\ \emph {et~al.}(2005)\citenamefont {Gubin}, \citenamefont {Il'in}, \citenamefont {Vitusevich}, \citenamefont {Siegel},\ and\ \citenamefont {Klein}}]{Gubin2005}%
  \BibitemOpen
  \bibfield  {author} {\bibinfo {author} {\bibfnamefont {A.~I.}\ \bibnamefont {Gubin}}, \bibinfo {author} {\bibfnamefont {K.~S.}\ \bibnamefont {Il'in}}, \bibinfo {author} {\bibfnamefont {S.~A.}\ \bibnamefont {Vitusevich}}, \bibinfo {author} {\bibfnamefont {M.}~\bibnamefont {Siegel}},\ and\ \bibinfo {author} {\bibfnamefont {N.}~\bibnamefont {Klein}},\ }\bibfield  {title} {\bibinfo {title} {Dependence of magnetic penetration depth on the thickness of superconducting {Nb} thin films},\ }\href {https://doi.org/10.1103/PhysRevB.72.064503} {\bibfield  {journal} {\bibinfo  {journal} {Phys. Rev. B}\ }\textbf {\bibinfo {volume} {72}},\ \bibinfo {pages} {064503} (\bibinfo {year} {2005})}\BibitemShut {NoStop}%
\bibitem [{\citenamefont {Brandt}(1994{\natexlab{a}})}]{Brandt1994a}%
  \BibitemOpen
  \bibfield  {author} {\bibinfo {author} {\bibfnamefont {E.~H.}\ \bibnamefont {Brandt}},\ }\bibfield  {title} {\bibinfo {title} {{Thin superconductors in a perpendicular magnetic ac field: General formulation and strip geometry}},\ }\href {https://doi.org/10.1103/PhysRevB.49.9024} {\bibfield  {journal} {\bibinfo  {journal} {Phys. Rev. B}\ }\textbf {\bibinfo {volume} {49}},\ \bibinfo {pages} {9024} (\bibinfo {year} {1994}{\natexlab{a}})}\BibitemShut {NoStop}%
\bibitem [{\citenamefont {Brandt}(1994{\natexlab{b}})}]{Brandt1994b}%
  \BibitemOpen
  \bibfield  {author} {\bibinfo {author} {\bibfnamefont {E.~H.}\ \bibnamefont {Brandt}},\ }\bibfield  {title} {\bibinfo {title} {{Thin superconductors in a perpendicular magnetic ac field. II. Circular disk}},\ }\href {https://doi.org/10.1103/PhysRevB.50.4034} {\bibfield  {journal} {\bibinfo  {journal} {Phys. Rev. B}\ }\textbf {\bibinfo {volume} {50}},\ \bibinfo {pages} {4034} (\bibinfo {year} {1994}{\natexlab{b}})}\BibitemShut {NoStop}%
\bibitem [{\citenamefont {Krawczyk}\ \emph {et~al.}(2012)\citenamefont {Krawczyk}, \citenamefont {Sokolovskyy}, \citenamefont {Klos},\ and\ \citenamefont {Mamica}}]{Krawczyk_2012}%
  \BibitemOpen
  \bibfield  {author} {\bibinfo {author} {\bibfnamefont {M.}~\bibnamefont {Krawczyk}}, \bibinfo {author} {\bibfnamefont {M.~L.}\ \bibnamefont {Sokolovskyy}}, \bibinfo {author} {\bibfnamefont {J.~W.}\ \bibnamefont {Klos}},\ and\ \bibinfo {author} {\bibfnamefont {S.}~\bibnamefont {Mamica}},\ }\bibfield  {title} {\bibinfo {title} {On the formulation of the exchange field in the {Landau-Lifshitz} equation for spin-wave calculation in magnonic crystals},\ }\href {https://doi.org/10.1155/2012/764783} {\bibfield  {journal} {\bibinfo  {journal} {Adv. Condens. Matter Phys.}\ }\textbf {\bibinfo {volume} {2012}},\ \bibinfo {pages} {764783} (\bibinfo {year} {2012})}\BibitemShut {NoStop}%
\bibitem [{\citenamefont {Tinkham}(2004)}]{Tinkham_2004}%
  \BibitemOpen
  \bibfield  {author} {\bibinfo {author} {\bibfnamefont {M.}~\bibnamefont {Tinkham}},\ }\href@noop {} {\emph {\bibinfo {title} {Introduction to Superconductivity}}}\ (\bibinfo  {publisher} {Dover Publications},\ \bibinfo {address} {New York},\ \bibinfo {year} {2004})\BibitemShut {NoStop}%
\bibitem [{\citenamefont {Tacchi}\ \emph {et~al.}(2012)\citenamefont {Tacchi}, \citenamefont {Duerr}, \citenamefont {Klos}, \citenamefont {Madami}, \citenamefont {Neusser}, \citenamefont {Gubbiotti}, \citenamefont {Carlotti}, \citenamefont {Krawczyk},\ and\ \citenamefont {Grundler}}]{Tacchi2012}%
  \BibitemOpen
  \bibfield  {author} {\bibinfo {author} {\bibfnamefont {S.}~\bibnamefont {Tacchi}}, \bibinfo {author} {\bibfnamefont {G.}~\bibnamefont {Duerr}}, \bibinfo {author} {\bibfnamefont {J.~W.}\ \bibnamefont {Klos}}, \bibinfo {author} {\bibfnamefont {M.}~\bibnamefont {Madami}}, \bibinfo {author} {\bibfnamefont {S.}~\bibnamefont {Neusser}}, \bibinfo {author} {\bibfnamefont {G.}~\bibnamefont {Gubbiotti}}, \bibinfo {author} {\bibfnamefont {G.}~\bibnamefont {Carlotti}}, \bibinfo {author} {\bibfnamefont {M.}~\bibnamefont {Krawczyk}},\ and\ \bibinfo {author} {\bibfnamefont {D.}~\bibnamefont {Grundler}},\ }\bibfield  {title} {\bibinfo {title} {Forbidden band gaps in the spin-wave spectrum of a two-dimensional bicomponent magnonic crystal},\ }\href {https://doi.org/10.1103/PhysRevLett.109.137202} {\bibfield  {journal} {\bibinfo  {journal} {Phys. Rev. Lett.}\ }\textbf {\bibinfo {volume} {109}},\ \bibinfo {pages} {137202} (\bibinfo {year} {2012})}\BibitemShut {NoStop}%
\bibitem [{\citenamefont {Doria}\ \emph {et~al.}(1989)\citenamefont {Doria}, \citenamefont {Gubernatis},\ and\ \citenamefont {Rainer}}]{Doria_1989}%
  \BibitemOpen
  \bibfield  {author} {\bibinfo {author} {\bibfnamefont {M.~M.}\ \bibnamefont {Doria}}, \bibinfo {author} {\bibfnamefont {J.~E.}\ \bibnamefont {Gubernatis}},\ and\ \bibinfo {author} {\bibfnamefont {D.}~\bibnamefont {Rainer}},\ }\bibfield  {title} {\bibinfo {title} {{Virial theorem for Ginzburg-Landau theories with potential applications to numerical studies of type-II superconductors}},\ }\href {https://doi.org/10.1103/PhysRevB.39.9573} {\bibfield  {journal} {\bibinfo  {journal} {Phys. Rev. B}\ }\textbf {\bibinfo {volume} {39}},\ \bibinfo {pages} {9573} (\bibinfo {year} {1989})}\BibitemShut {NoStop}%
\bibitem [{\citenamefont {Il’in}\ \emph {et~al.}(2010)\citenamefont {Il’in}, \citenamefont {Rall}, \citenamefont {Siegel}, \citenamefont {Engel}, \citenamefont {Schilling}, \citenamefont {Semenov},\ and\ \citenamefont {Huebers}}]{Ilin_2020}%
  \BibitemOpen
  \bibfield  {author} {\bibinfo {author} {\bibfnamefont {K.}~\bibnamefont {Il’in}}, \bibinfo {author} {\bibfnamefont {D.}~\bibnamefont {Rall}}, \bibinfo {author} {\bibfnamefont {M.}~\bibnamefont {Siegel}}, \bibinfo {author} {\bibfnamefont {A.}~\bibnamefont {Engel}}, \bibinfo {author} {\bibfnamefont {A.}~\bibnamefont {Schilling}}, \bibinfo {author} {\bibfnamefont {A.}~\bibnamefont {Semenov}},\ and\ \bibinfo {author} {\bibfnamefont {H.-W.}\ \bibnamefont {Huebers}},\ }\bibfield  {title} {\bibinfo {title} {Influence of thickness, width and temperature on critical current density of nb thin film structures},\ }\href {https://doi.org/https://doi.org/10.1016/j.physc.2010.02.042} {\bibfield  {journal} {\bibinfo  {journal} {Physica C}\ }\textbf {\bibinfo {volume} {470}},\ \bibinfo {pages} {953} (\bibinfo {year} {2010})},\ \bibinfo {note} {vortex Matter in Nanostructured Superconductors}\BibitemShut {NoStop}%
\bibitem [{\citenamefont {Gurevich}\ and\ \citenamefont {Melkov}(1996)}]{Gurevich1996}%
  \BibitemOpen
  \bibfield  {author} {\bibinfo {author} {\bibfnamefont {A.}~\bibnamefont {Gurevich}}\ and\ \bibinfo {author} {\bibfnamefont {G.}~\bibnamefont {Melkov}},\ }\href@noop {} {\emph {\bibinfo {title} {Magnetization oscillations and waves}}}\ (\bibinfo  {publisher} {CRC Press},\ \bibinfo {address} {London},\ \bibinfo {year} {1996})\BibitemShut {NoStop}%
\bibitem [{\citenamefont {Kaczér}\ and\ \citenamefont {Murtinová}(1974)}]{Kaczer_1974}%
  \BibitemOpen
  \bibfield  {author} {\bibinfo {author} {\bibfnamefont {J.}~\bibnamefont {Kaczér}}\ and\ \bibinfo {author} {\bibfnamefont {L.}~\bibnamefont {Murtinová}},\ }\bibfield  {title} {\bibinfo {title} {On the demagnetizing energy of periodic magnetic distributions},\ }\href {https://doi.org/https://doi.org/10.1002/pssa.2210230108} {\bibfield  {journal} {\bibinfo  {journal} {Phys. Status Solidi A}\ }\textbf {\bibinfo {volume} {23}},\ \bibinfo {pages} {79} (\bibinfo {year} {1974})}\BibitemShut {NoStop}%
\bibitem [{\citenamefont {Yariv}\ and\ \citenamefont {P.~Yeh}(2003)}]{Yariv2003}%
  \BibitemOpen
  \bibfield  {author} {\bibinfo {author} {\bibfnamefont {A.}~\bibnamefont {Yariv}}\ and\ \bibinfo {author} {\bibfnamefont {O.}~\bibnamefont {P.~Yeh}},\ }\href@noop {} {\emph {\bibinfo {title} {Optical Waves in Crystals}}}\ (\bibinfo  {publisher} {Wiley- Interscience},\ \bibinfo {address} {New York},\ \bibinfo {year} {2003})\ p.\ \bibinfo {pages} {161}\BibitemShut {NoStop}%
\bibitem [{\citenamefont {Mieszczak}\ and\ \citenamefont {Kłos}(2022)}]{mieszczak_2022}%
  \BibitemOpen
  \bibfield  {author} {\bibinfo {author} {\bibfnamefont {S.}~\bibnamefont {Mieszczak}}\ and\ \bibinfo {author} {\bibfnamefont {J.~W.}\ \bibnamefont {Kłos}},\ }\bibfield  {title} {\bibinfo {title} {Interface modes in planar one-dimensional magnonic crystals},\ }\href {https://doi.org/10.1038/s41598-022-15328-x} {\bibfield  {journal} {\bibinfo  {journal} {Sci. Rep.}\ }\textbf {\bibinfo {volume} {12}},\ \bibinfo {pages} {11335} (\bibinfo {year} {2022})}\BibitemShut {NoStop}%
\bibitem [{\citenamefont {Mruczkiewicz}\ \emph {et~al.}(2013)\citenamefont {Mruczkiewicz}, \citenamefont {Krawczyk}, \citenamefont {Gubbiotti}, \citenamefont {Tacchi}, \citenamefont {Filimonov}, \citenamefont {Kalyabin}, \citenamefont {Lisenkov},\ and\ \citenamefont {Nikitov}}]{mruczkiewicz_2013}%
  \BibitemOpen
  \bibfield  {author} {\bibinfo {author} {\bibfnamefont {M.}~\bibnamefont {Mruczkiewicz}}, \bibinfo {author} {\bibfnamefont {M.}~\bibnamefont {Krawczyk}}, \bibinfo {author} {\bibfnamefont {G.}~\bibnamefont {Gubbiotti}}, \bibinfo {author} {\bibfnamefont {S.}~\bibnamefont {Tacchi}}, \bibinfo {author} {\bibfnamefont {Y.~A.}\ \bibnamefont {Filimonov}}, \bibinfo {author} {\bibfnamefont {D.~V.}\ \bibnamefont {Kalyabin}}, \bibinfo {author} {\bibfnamefont {I.~V.}\ \bibnamefont {Lisenkov}},\ and\ \bibinfo {author} {\bibfnamefont {S.~A.}\ \bibnamefont {Nikitov}},\ }\bibfield  {title} {\bibinfo {title} {Nonreciprocity of spin waves in metallized magnonic crystal},\ }\href {https://doi.org/10.1088/1367-2630/15/11/113023} {\bibfield  {journal} {\bibinfo  {journal} {New J. Phys.}\ }\textbf {\bibinfo {volume} {15}},\ \bibinfo {pages} {113023} (\bibinfo {year} {2013})}\BibitemShut {NoStop}%
\bibitem [{\citenamefont {Lisenkov}\ \emph {et~al.}(2015)\citenamefont {Lisenkov}, \citenamefont {Kalyabin}, \citenamefont {Osokin}, \citenamefont {Klos}, \citenamefont {Krawczyk},\ and\ \citenamefont {Nikitov}}]{lisenkov_2015}%
  \BibitemOpen
  \bibfield  {author} {\bibinfo {author} {\bibfnamefont {I.}~\bibnamefont {Lisenkov}}, \bibinfo {author} {\bibfnamefont {D.}~\bibnamefont {Kalyabin}}, \bibinfo {author} {\bibfnamefont {S.}~\bibnamefont {Osokin}}, \bibinfo {author} {\bibfnamefont {J.}~\bibnamefont {Klos}}, \bibinfo {author} {\bibfnamefont {M.}~\bibnamefont {Krawczyk}},\ and\ \bibinfo {author} {\bibfnamefont {S.}~\bibnamefont {Nikitov}},\ }\bibfield  {title} {\bibinfo {title} {{Nonreciprocity of edge modes in 1D magnonic crystal}},\ }\href {https://doi.org/https://doi.org/10.1016/j.jmmm.2014.10.073} {\bibfield  {journal} {\bibinfo  {journal} {J. Magn. Magn. Mater.}\ }\textbf {\bibinfo {volume} {378}},\ \bibinfo {pages} {313} (\bibinfo {year} {2015})}\BibitemShut {NoStop}%
\bibitem [{\citenamefont {Zhou}\ \emph {et~al.}(2024)\citenamefont {Zhou}, \citenamefont {Ye}, \citenamefont {Bai},\ and\ \citenamefont {Yu}}]{zhou_2024}%
  \BibitemOpen
  \bibfield  {author} {\bibinfo {author} {\bibfnamefont {X.-H.}\ \bibnamefont {Zhou}}, \bibinfo {author} {\bibfnamefont {X.}~\bibnamefont {Ye}}, \bibinfo {author} {\bibfnamefont {L.}~\bibnamefont {Bai}},\ and\ \bibinfo {author} {\bibfnamefont {T.}~\bibnamefont {Yu}},\ }\bibfield  {title} {\bibinfo {title} {Giant enhancement of magnon transport by superconductor meissner screening},\ }\href {https://doi.org/10.1103/PhysRevB.110.L020404} {\bibfield  {journal} {\bibinfo  {journal} {Phys. Rev. B}\ }\textbf {\bibinfo {volume} {110}},\ \bibinfo {pages} {L020404} (\bibinfo {year} {2024})}\BibitemShut {NoStop}%
\end{thebibliography}

%

%

\clearpage

\section*{supplementary information}
\renewcommand{\thesubsection}{SI \arabic{subsection}}
\subsection{Plane-wave method}\label{app:PWM}

In our studies, we do not consider the perpendicular standing SW modes, which have much higher frequencies. Therefore, for thin FM films, the SW amplitude can be assumed to be uniform across the thickness of the film: $m_{k,\alpha}(x,y)\approx m_{k,\alpha}(x)$, where $\alpha=x,z$. 

The amplitude of SWs propagating in MC has a form of Bloch functions: $m_{k,\alpha}(x)=u_{k,\alpha}(x)e^{i k x}$. The periodic component of the Bloch function $u_{k,\alpha}(x)$ can be expanded in the Fourier series, and then the Bloch function can be written as
\begin{equation}
    m_{k,\alpha}(x)=\sum_{G}m_{k,\alpha,G}e^{i(k+G)x},
     \label{eq:m_Fourier}
\end{equation}
where $G=2\pi n/a$ is a reciprocal-lattice vector, indexed by integer $n=0,\pm 1,\pm 2,\ldots$. The symbols $m_{\alpha,k,G}$ are the coefficients of the Fourier series of $u_{k,\alpha}(x)$. The stray field $H_{\rm sc}(x)$, as a periodic function, can also be expanded in the Fourier series:
\begin{equation}
    H_{\rm sc}(x)=\sum_{G}H_{{\rm sc},G}e^{iGx},
     \label{eq:Hsc_Fourier}
\end{equation}
where $H_{{\rm sc},G}$ are the coefficients of this expansion.

Dynamic demagnetizing field $\mathbf{h}_{\rm{d}}(x,y)$ is dependent on the spatial distribution of dynamic magnetization Eq.~(7). For planar MC, the demagnetizing field can be expressed in terms of the Fourier coefficients for magnetization distribution \cite{Kaczer_1974}. Adopting the approach of \cite{Kaczer_1974} to a dynamical case, we obtain the following relations:

\begin{equation}
    \begin{split}
    h_{{\rm d},x}(x,y)&=-\sum_{G}m_{k,x,G}e^{i(G+k)x} A(y,G+k),\\
    h_{{\rm d},z}(x,y)&=0.
     \end{split}
     \label{eq:h_demag}
\end{equation}

The function $A(y,\kappa)$, which appears in (\ref{eq:h_demag}), has the form:
\begin{equation}
    A(y,\kappa)=1-\frac{\cosh(|\kappa| (y-y_0))}{\cosh\left(\frac{|\kappa|d}{2}\right)+\sinh\left(\frac{|\kappa|d}{2}\right)},
\end{equation}
where $y_0=s-\tfrac{t+d}{2}$ is the center of FM layer.
Since the FM layer is thin and $h_{{\rm d},x}(x,y)$ does not change significantly across the film thickness, we can take its value from the center of the layer, i.e. assume that $h_{{\rm d},x}(x)=h_{d,x}(x,y=y_0)$.

Substituting the Fourier expansions (\ref{eq:m_Fourier}-\ref{eq:h_demag}) into Eq.~(8) leads to algebraic eigenvalue problem Eq.~(9) where the elements of $\mathbf{\bar{\bar{M}}}_{xz}$ and $\mathbf{\bar{\bar{M}}}_{zx}$ have the following form:

\begin{equation}
    \begin{split}
        M_{xz,G,G'}=&-\delta_{G,G'}\left(1+\lambda_{\rm ex}^2(G+k)^2\frac{M_{\rm s}}{\Tilde{H}_0}\right)
        -\frac{H_{{\rm sc},G'-G}}{\Tilde{H}_0},\\
        M_{zx,G,G'}=\;\;&\delta_{G,G'}\left(1+\left(\lambda_{\rm ex}^2(G+k)^2+A(y_0,G+k)\right)\frac{M_{\rm s}}{\Tilde{H}_0}\right)\\
        +\frac{H_{{\rm sc},G'-G}}{\Tilde{H}_0},
        \label{eq:Matrix_el}
   \end{split}
\end{equation}
where $\delta_{G,G'}$ is the Kronecker delta.

To solve the eigenproblem Eq.~(9) numerically, we have to approximate the infinite Fourier expansions by finite ones, i.e. constrain the range of the reciprocal-lattice vectors $G=2\pi n/a$ to $n=0,\pm 1,\pm 2,\ldots, N$. For considered system, the range $N=15$ gives satisfactory results for about six lowest frequency bands. 

\subsection{Width of the band gaps -- dependence on B$_{0}$ and gaps between SC strips}\label{app:Gand_gap}

\begin{figure}[h!]
\centering
    \includegraphics[width=0.95\columnwidth]{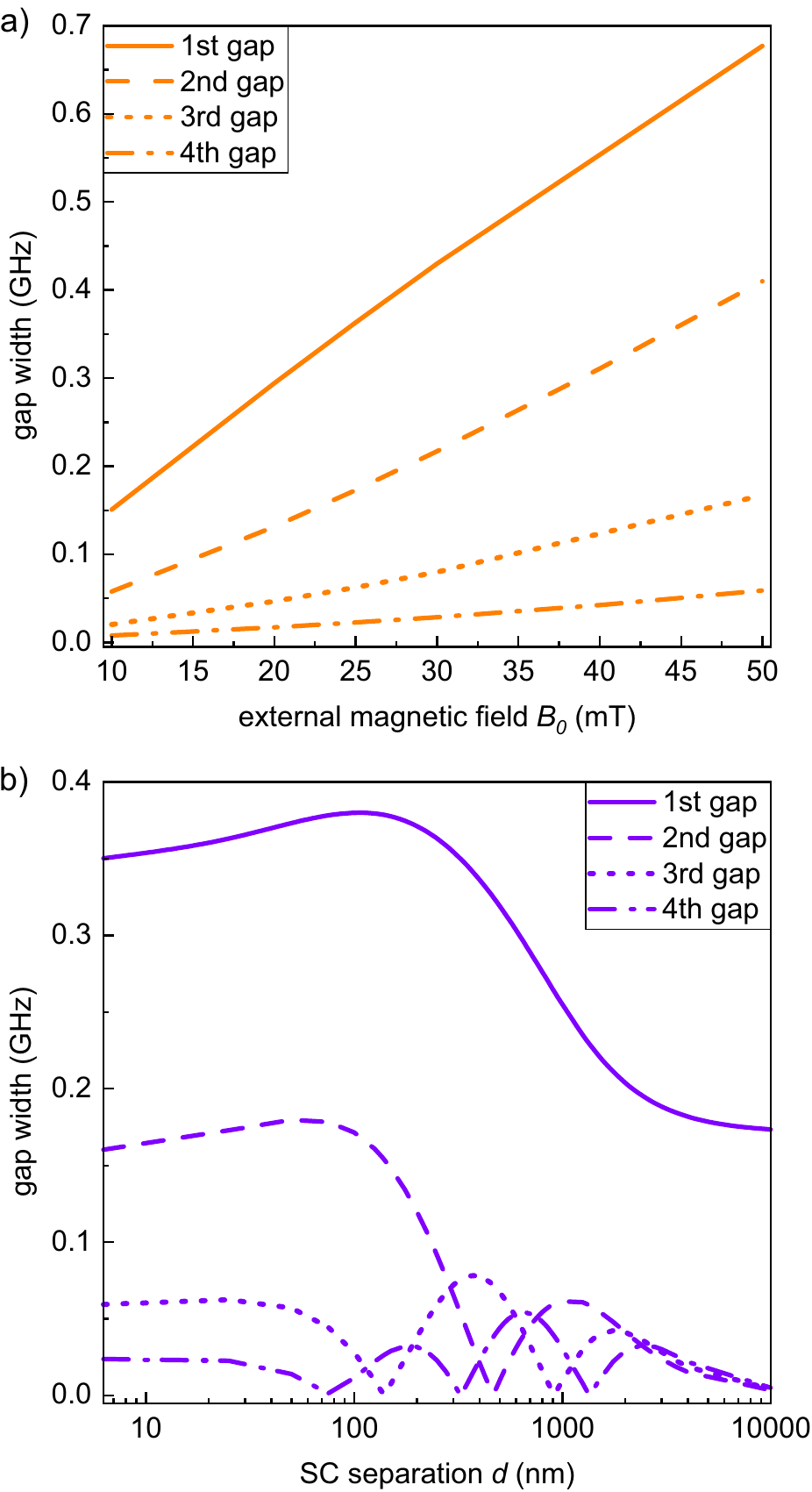}
    \caption{
    The width of the forbidden frequency gaps as a function of (a) external magnetic field and (b) superconducting strip spacing are shown. Parts of the drawing (a,b) correspond to Fig.~6(a) and Fig.~6(b), respectively. The orange (violet) numbers correspond to the gap number marked on Fig.~6(a) (Fig.~6(b)). On (b), the calculations have been performed for a separation range from 6.25~nm to 10~µm.
    }
\label{fig:appx_freq_gap}
\end{figure}

In the discussion of Fig.~6, we briefly described how the width of the frequency gaps is modified with the external field $B_0$ and the distance between the SC strips $d$. To demonstrate the mentioned almost linear increase of the gap width with $B_0$ and the multiple gaps closing with $d$, we prepared Fig.~\ref{fig:appx_freq_gap}.




The dependence of the width of the frequency gaps on the applied field, shown in Fig.~\ref{fig:appx_freq_gap}(a), is close to linear. This is due to the fact that the profile of the stray field $\mathbf{B}_{\rm sc}(x)$ scales linearly with the external field. This means that all of its Fourier components also scale linearly with $B_0$. Since the field $\mathbf{B}_{\rm sc}(x)$ plays the role of a periodic coefficient in the linearized LL equations describing the SW dynamics in the considered MC, then the width of successive frequency gaps expressed by the corresponding Fourier coefficients should also have a linear dependence on $B_0$ \cite{Tacchi2012, Yariv2003}. The linear scalability of the frequency gaps with an external field is a convenient feature of the MCs proposed in this work.

Fig.~\ref{fig:appx_freq_gap}(b) presents the dependence of the width of successive gaps on the separation between SC strips. In the limit $d\rightarrow\infty$, the SW spectrum of MC is the same as the spectrum for the isolated well of the stray field. The magnonic bands for the frequencies below (above) the FMR frequency of the pristine FM layer will merge into a single level corresponding to bound states in the well (to continuous spectrum without gaps). Therefore in the limit $d\rightarrow\infty$, the width of the first gap, which separates the band laying below the FMR frequency, has finite width, while the higher gaps gradually disappear. This effect is clearly visible in Fig.~\ref{fig:appx_freq_gap}(b).   


On the other hand, for small separation between SC stripes $d$, we obtain unintuitive results. Even for small $d$, when the SC system seems to be almost continuous, the frequency gap stay wide open. It can be understood when we recall (see Fig.~2) that barriers of stray field become not only narrower with decreasing $d$ but also higher. The competition between the width and height of the barrier determines the strength of the SW scattering the the width of frequency gaps.




In Fig.~\ref{fig:appx_freq_gap}(b), it is easier to notice that the $n$th gap is closed $n-1$ times when bulk parameter $d$ is swept over its whole range. It is explained by the effect of band crossing \cite{mieszczak_2022}, known for wave excitations in periodic structures of different kinds, e.g. in photonic and phononic crystals.

\subsection{Dynamical coupling between ferrimagnetic layer and superconducting pattern -- FEM study}\label{app:FEM_dynamics}

\begin{figure}[t!]
\centering
    \includegraphics[width=0.95\columnwidth]{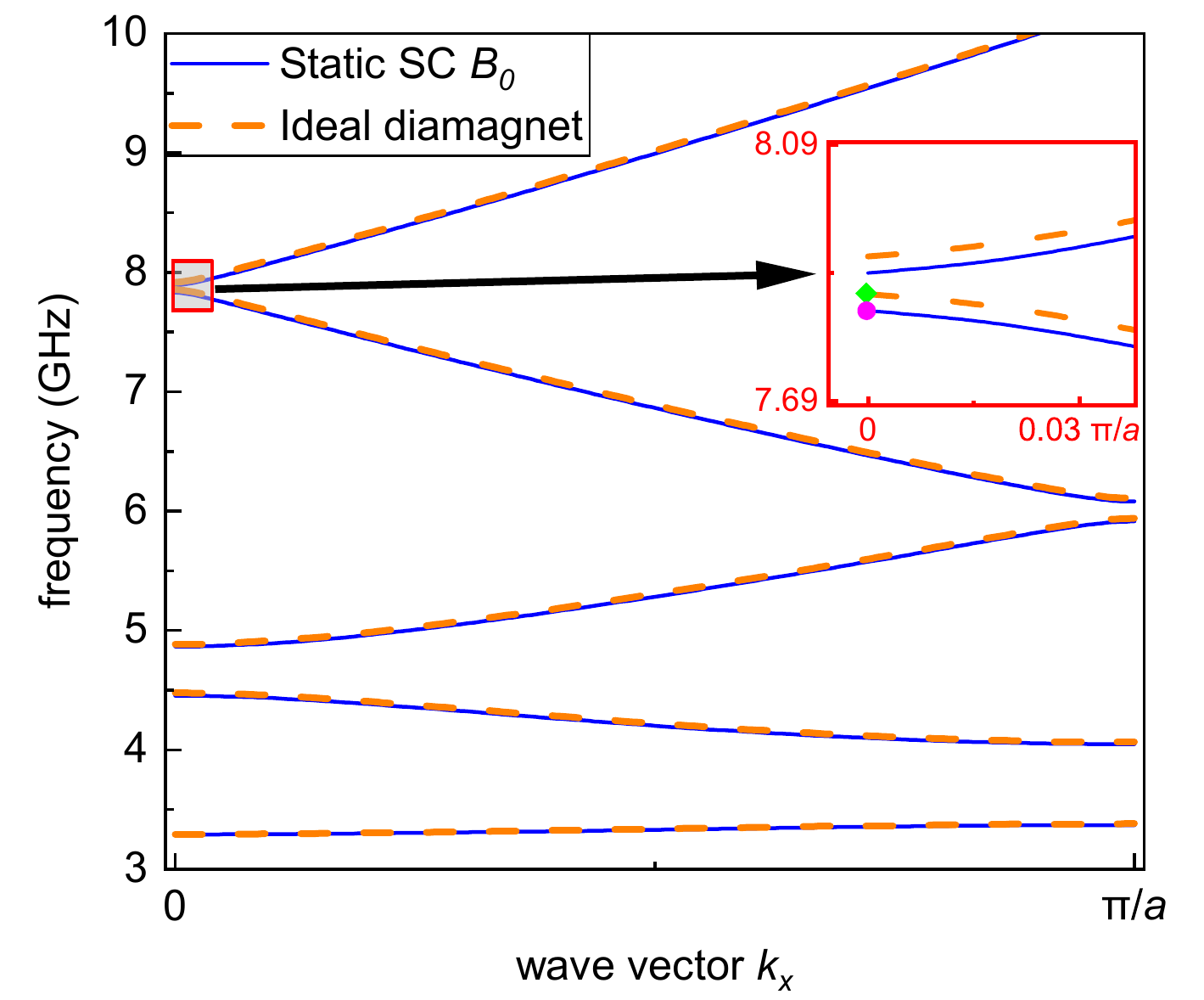}
    \caption{ The FEM calculations of the SW dispersion relation for $B_0=50$~mT, $d=25$~nm, $w=400$~nm, without (blue lines) and with (orange lines) dynamic coupling between SC strips and FM layer. The coupling was introduced under the assumption of ideal diamagnetism of SC material where the field was completely expelled from the SC strips. The inset shows a tiny difference in the SW frequencies for both considered approaches. The magenta and green symbols mark the modes for which the maps of the dynamic demagnetizing field have been plotted in Fig.~\ref{fig:hx_profiles}.
    }
\label{fig:appx_ideal_diamagnet}
\end{figure}

\begin{figure}[t!]
\centering
    \includegraphics[width=0.95\columnwidth]{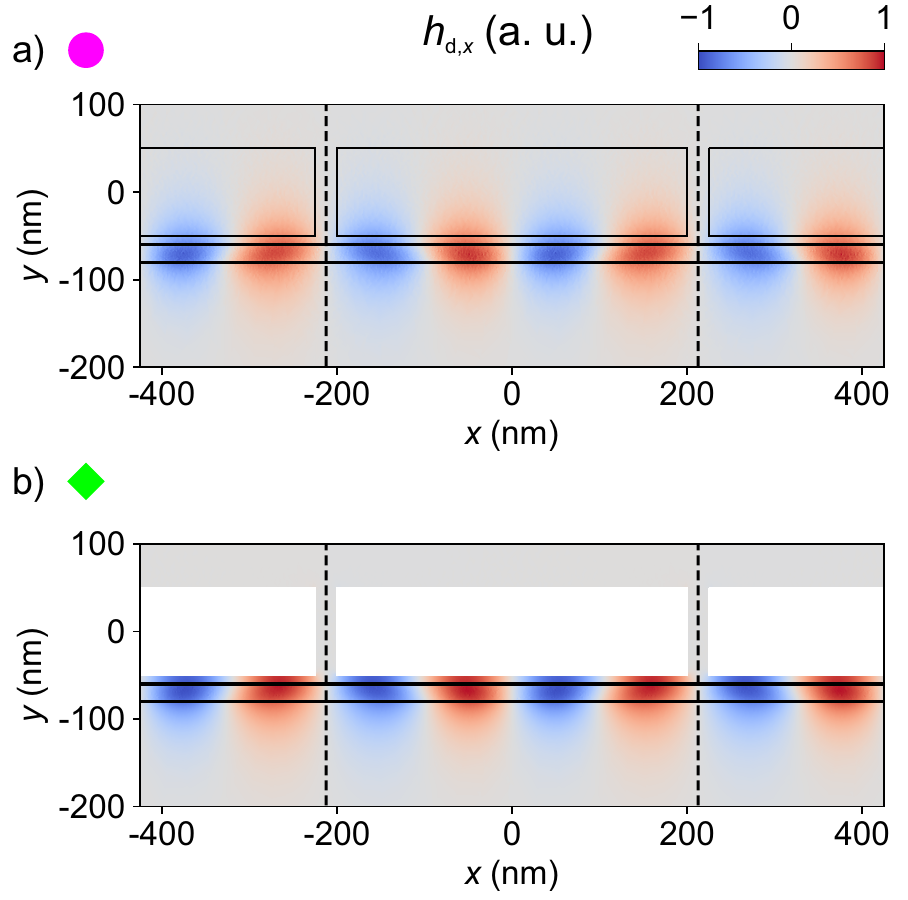}
    \caption{
  The maps of the real part of the in-plane component of the dynamic demagnetizing field $\Re[h_{\mathrm{d},x}(x,y)]$, for the mode marked by magenta and green in the inset of Fig.~\ref{fig:appx_ideal_diamagnet} -- fourth mode at $k_x=0$ in (a) absence and (b) presence of the dynamic coupling with SC strips approximated as ideal diamagnets. In both cases, the static component of the field produced by the SC strips is included. Two horizontal lines indicate the region of the FM layer. The black rectangles in (a) denote the domains of the SC strips, which expel only the static field but are transparent to the dynamic demagnetizing field produced by the propagating SWs. The white rectangles in (b) represent the regions of the SC strips where both static and dynamic components of the field are shielded.
    }
\label{fig:hx_profiles}
\end{figure}

The comparative FEM calculations, demonstrating the impact of dynamic coupling between SC strips and FM layer, were performed in COMSOL Multiphysics in the following way. Initially, we solved the London equation assuming only the presence of the SC strip to calculate the static magnetic field produced by the superconductor in the absence of a ferromagnet. In these calculations, we assumed the finite London penetration depth $\lambda=50$~nm. We considered the large vacuum domain above and below the considered system which ensures that the field distribution will be properly calculated and Bloch boundary conditions on the edges of unit cell -- dashed lines in Fig.~\ref{fig:hx_profiles}. Then, we solved the LL equation and Gauss equation for magnetism within the magnetostatic approximation \cite{Kharlan2024} to determine the SW dynamics, taking into account the static field produced by the SC strip calculated in the previous step. In this step, we performed two studies: (i) the SC strip was treated as a vacuum and did not produce any dynamic magnetic field, (ii) the SC strip was replaced with an ideal diamagnet for dynamic demagnetizing fields. The implementation of an ideal diamagnet is based on the boundary condition for the zeroing of the normal component of the magnetic field so that the magnetic field cannot penetrate the diamagnet. 
 In COMSOL, this kind of boundary conditions are applied automatically by removing the volume of SC from the computational domain (see the white areas in Fig.~\ref{fig:hx_profiles}(b)).

Fig.~\ref{fig:appx_ideal_diamagnet} shows that the dynamic coupling is very small for considered configuration, i.e. for the field $B_0$ applied at normal direction to the FM layer and SC strips.The dispersion branches for the hybrid system where only the static coupling was induced (blue line) are shifted down in frequency by a negligible amount when compared to the results where both the static coupling and the shielding of the dynamic demagnetization field were considered (orange lines).

To check that the effect of an ideal diamagnet was calculated properly, we show the profiles of $h_x$ dynamic field produced by the fourth localized mode in the case without diamagnet in Fig.~\ref{fig:hx_profiles}(a), and with diamagnet in Fig.~\ref{fig:hx_profiles}(b). When the diamagnet is absent, the dynamic field penetrates the area where the SC should be present. When the diamagnet is present, the dynamic field is repelled from the diamagnet area, and its value increases inside the FM layer. 

The mechanism of dynamic shielding of the magnetic field generated by SWs is well-known effect \cite{Gurevich1996,mruczkiewicz_2013,lisenkov_2015} in the case of conventional conductors. 
It is worth noting that  dynamic shielding by superconductor in the Meissner state is more complicated \cite{borst_2023,zhou_2024} than static one, which is manifestation of ideal diamagnetism. The application of the $\mathbf{m}_{\rm eff}(\mathbf{r},t)=-\mathbf{h}(\mathbf{r},t)$ relation valid for the static case in the $\lambda\rightarrow 0$ limit is, in general, not correct for dynamic shielding by superconductor. The relation between dynamic field, on which the superconductor is exposed $\mathbf{h}(\mathbf{r})e^{i\omega t}$, and its response, described by effective magnetization $\mathbf{m}_{\rm eff}(\mathbf{r})e^{i\omega t}$, is given by the formula: \cite{borst_2023,zhou_2024}
\begin{equation}
    \frac{1}{\lambda^2}\mathbf{h}(\mathbf{r})=\Delta \mathbf{m}_{\rm eff}(\mathbf{r})\label{eq:dyn_shield}.
\end{equation}
 Eq.~\ref{eq:dyn_shield} is derived from Faraday's law: $\nabla\!\times\!\mathbf{j}(\mathbf{r},t)\!=\!-\sigma/\mu_0\;\partial_t\mathbf{h}(\mathbf{r},t)$ and the relation between (superconducting) current $\mathbf{j}(\mathbf{r},t)=\mathbf{j}(\mathbf{r})e^{i\omega t}$ and (effective) magnetization: $\mathbf{j}(\mathbf{r},t)=\nabla\times\mathbf{m}_{\rm eff}(\mathbf{r},t)$, taking the conductivity in the form typical for superconductors: $\sigma=i\,1/(\omega \mu_0 \lambda^2)$. It is reasonable to assume that the field produced by the SW (see Fig.~5) and the resulting profile of the effective magnetization in the superconductor have the wavy pattern. Its wavelength $\lambda_{m_{\rm eff}}$ can be roughly estimated from  Fig.~5. For example, for mode No.~3: $\lambda_{m_{\rm eff}}\approx 400$~nm and for modes Nos.~7 and 8, marked also in Fig.~\ref{fig:appx_ideal_diamagnet}: $\lambda_{m_{\rm eff}}\approx 200$~nm. Under this assumption, we estimate from Eq.~\ref{eq:dyn_shield} the strength of dynamic shielding:
 \begin{equation}
     \mathbf{m}_{\rm eff}(\mathbf{r})=-\frac{1}{(2\pi)^2}\left(\frac{\lambda_{m_{\rm eff}}}{\lambda}\right)^2\mathbf{h}(\mathbf{r}).\label{eq:dyn_shield_app}
 \end{equation}
The factor $-1/(2\pi)^2(\lambda_{m_{\rm eff}}/\lambda)^2$ takes the values about $-1.6$ and $-0.4$ for the mods No.~3 and Nos.~7 and 8, respectively. The FEM simulation were performed for the fixed value of this factor equal to $-1$. We think that even an enhancement according to Eq.~\ref{eq:dyn_shield_app} will not produce the noticeable change in the dispersion relation in Fig.~\ref{fig:appx_ideal_diamagnet}.

\end{document}